\documentclass[pre,twocolumn]{revtex4-2}

\usepackage{graphicx}
\usepackage{amsmath,mathtools,amssymb}
\usepackage{bm}
\usepackage{hyperref}

\usepackage[utf8]{inputenc}
\usepackage[T1]{fontenc}
\usepackage{mathptmx}
\usepackage{bm}
\usepackage{pdfrender}
\usepackage{physics}
\usepackage{cases}

\usepackage{amsmath}
\usepackage{mathtools}
\usepackage{amssymb}
\usepackage{tikz}

\usepackage{pst-node}
\usepackage{verbatim}   
\usepackage{color}      

\usepackage{braket}

\newcommand{\ba}{\begin{eqnarray}}
\newcommand{\ea}{\end{eqnarray}}
\newcommand{\be}{\begin{equation}}
\newcommand{\ee}{\end{equation}}

\begin{document}

\title{A new representation of effective diffusion in fluctuating periodic potentials through stochastic duality}

\author{Derek Frydel}
\affiliation{Department of Chemistry, Universidad Técnica Federico Santa María, Campus San Joaquín, 7820275, Santiago, Chile}

\date{\today}

\begin{abstract}
We revisit the classical problem of effective diffusion in fluctuating periodic force fields by combining homogenization with stochastic duality. 
We consider a specific class of fluctuations in which the internal state of the force field is independently redrawn from a fixed 
distribution at Poisson-distributed times. In the homogenization framework, the effective diffusion coefficient \(D_{\rm eff}\) is 
expressed in terms of a corrector function satisfying a backward Kolmogorov equation. For the redraw dynamics considered 
here, the equation governing the spatial derivative of the corrector can be reinterpreted as the stationary forward equation of a 
dual system obtained by reversing the force field, \(F\to-F\). This relation eliminates the corrector and yields a representation 
of \(D_{\rm eff}\) entirely in terms of stationary densities and probability currents of the physical and dual systems. We further 
introduce an auxiliary function \(Q(x)\) whose spatial variation is determined by the stationary current. When this current vanishes, 
\(Q(x)\) becomes independent of position and the effective diffusion coefficient reduces to the local relation
\(
D_{\rm eff}=L^2Q.
\)
This result generalizes the classical Lifson--Jackson relation to fluctuating periodic force fields. We verify the theory using 
analytical calculations, numerical solutions of the stationary equations, and direct simulations of the long-time particle 
displacement, including a driven system with nonzero stationary current. 
\end{abstract}

\pacs{
}

\maketitle

\section{Introduction}

Diffusion in periodic potentials is a classical problem in stochastic dynamics. On length scales much larger than the spatial period, 
the motion of a Brownian particle is characterized by an effective drift velocity and an effective diffusion coefficient \(D_{\rm eff}\), 
which generally differs from the thermal diffusion coefficient \(D\). For a particle in an equilibrium periodic potential, 
\(D_{\rm eff}\) was obtained by Lifson and Jackson more than sixty years ago~\cite{JCP-Lifson-1962}. 
The problem was later extended to tilted periodic potentials, where the interplay between thermal fluctuations, periodic barriers, 
and a constant driving force can produce a strong enhancement of diffusion, commonly referred to as giant 
diffusion~\cite{PRL-Reimann-2001,PRE-Reimann-2002,Lindner-FNL-2001,PRE-Akimoto-2025}. 
For roughly three decades, related questions have also been studied for periodic potentials that fluctuate in 
time~\cite{AP-Abad-1998,PRE-Dubkov-2005,EPJB-Dubkov-2006,PA-Gleeson-2009}. 
Such fluctuations introduce an additional time scale and can substantially alter the long-time transport of the particle. 
They arise in models of stochastic ratchets and molecular motors
~\cite{PRL-Astumian-1994,PRL-Faucheux-1995,PR-Reimann-2002,RMP-Hanggi-2009}. 
Closely related transport problems occur for persistent random walks and active particles, including run-and-tumble particles, 
where stochastic changes in the direction or magnitude of the self-propulsion force constitute a fluctuating 
drive~\cite{PRL-Lindner-2008,SM-Koumakis-2014,EPL-Schehr-2020,EPJE-Su-2023,PRE-Roberts-2023}.

In this work, we revisit effective diffusion in periodic potentials by combining homogenization with stochastic duality. 
Homogenization is a coarse-graining procedure used to determine the effective macroscopic properties of systems 
with microscopic spatial structure~\cite{JSP-Latorre-2013,FNL-Pavliotis-2008,book-Bensoussan-1978,book-Pavliotis-2008}. 
It is particularly well suited to periodic potentials, where the repeated unit-cell structure gives rise to effective transport over 
much larger length scales.

We combine the homogenization framework with stochastic duality. A classical example is Siegmund duality, introduced in 
1976~\cite{AP-Siegmund-1976}, which relates first-passage properties of one stochastic process to probability distributions 
of a corresponding dual process. More recently, this idea has been applied to active particles and extended to other classes 
of stochastic processes~\cite{JPA-Gueneau-2024,POF-Frydel-2024,PRL-Mathis-2026,JCP-Frydel-2026,JSTAT-Gueneau-2024}. 
In the present work, stochastic duality provides a physical interpretation of the corrector function introduced by homogenization.

In this work, we focus on a specific type of fluctuations. We consider a fluctuating periodic force field,
\(
F(x,\sigma)=F(x+L,\sigma),
\)
whose internal state \(\sigma\) is independently redrawn from a fixed distribution \(p(\sigma)\) at Poisson-distributed times. 
The dual system for this type of internal-state dynamics is obtained by reversing the force, \(F\to-F\). 
Using the dual representation, we express \(D_{\rm eff}\) in terms of the stationary densities and probability currents of the physical 
and dual systems.

In any stationary system, the total probability current through a single cell is independent of position:
\[
j_{\rm ss}
=
\int d\sigma\,p(\sigma)j(x,\sigma)
=
\mathrm{const},
\]
where \(j(x,\sigma)\) is the spatial probability current associated with the internal state \(\sigma\).
This spatial invariance follows directly from stationarity. Our analysis reveals an additional position-independent quantity 
that emerges when the stationary current vanishes. This quantity, expressed in terms of stationary properties of the physical 
and dual systems, is
\begin{align}
Q(x)
&=
D\int d\sigma\,p(\sigma)\,n(x,\sigma)n_D(x,\sigma)
\nonumber\\
&\quad-
\tau\int d\sigma\,p(\sigma)\,j(x,\sigma)j_D(x,\sigma), \nonumber
\end{align}
where \(n\) and \(n_D\) are the stationary densities of the physical and dual systems, while \(j\) and \(j_D\) are their 
corresponding probability currents. We show that
\[
j_{\rm ss}=0
\qquad\Longrightarrow\qquad
Q(x)=\mathrm{const}.
\]
Thus, when the conserved stationary current vanishes, \(Q(x)\) becomes a second spatial invariant.

The significance of this result follows from the representation
\begin{align*}
D_{\rm eff}
&=
L^2\int_0^L dx\,\bar n(x)Q(x)
\\
&\quad+
L^2\tau j_{\rm ss}
\int_0^L dx\int d\sigma\,p(\sigma)\,
n(x,\sigma)j_D(x,\sigma).
\end{align*}
When \(j_{\rm ss}=0\), the second term vanishes and \(Q(x)=Q\) can be taken outside the first integral. Since the marginal 
distribution $\bar n(x)$ is normalized, we obtain
\[
D_{\rm eff}=L^2Q.
\]
The additional spatial invariant \(Q\) is therefore directly related to the effective diffusion coefficient. 
In this case, \(D_{\rm eff}\) admits a local representation. 
Since the 
above result recovers the Lifson--Jackson formula for a non-fluctuating system, which was also derived under the condition 
\(j_{\rm ss}=0\), we refer to this result as the generalized Lifson--Jackson formula.

In the second part of the paper, we apply the framework to several simple systems. The resulting formulas are tested against 
previously established results and against direct, more computationally expensive simulations of \(D_{\rm eff}\). Most of our 
attention is devoted to testing the generalized Lifson--Jackson formula, which applies when \(j_{\rm ss}=0\).

The paper is organized as follows. In Sec.~\ref{sec:model}, we define the model and introduce the stationary quantities 
and operators used throughout the paper. In Secs.~\ref{sec:homogenization} and \ref{sec:dual-system}, we develop 
the homogenization framework and relate the corrector function to the stationary distribution of the dual system. 
This relation is used in Sec.~\ref{sec:Deff} to express the effective diffusion coefficient in terms of stationary quantities 
of the physical and dual systems. In Sec.~\ref{sec:Q}, we introduce the auxiliary function \(Q(x)\) and derive an alternative 
representation of \(D_{\rm eff}\). In Sec.~\ref{sec:LJ}, we consider systems with zero stationary current and derive the 
generalized Lifson--Jackson relation. The symmetry relations between the physical and dual systems are established in 
Sec.~\ref{sec:symmetry}.

The applications begin in Sec.~\ref{sec:no-switching}, where we recover the known results for a non-fluctuating periodic 
force field. We then consider the two-state model with \(\sigma\in\{-1,1\}\) in Sec.~\ref{sec:2-state} and the on--off model 
with \(\sigma\in\{0,1\}\) in Sec.~\ref{sec:two-state-on-off}. For both models, we compare the analytical results with direct 
simulations and first-passage calculations. In Sec.~\ref{sec:finite-driving}, we apply the general dual representation to 
a fluctuating system with a finite driving force and nonzero stationary current. Finally, the main results are summarized 
in Sec.~\ref{sec:summary}.

\section{The model}
\label{sec:model}

This work considers the overdamped motion of a Brownian particle in a one-dimensional periodic force field that fluctuates in time. 
To incorporate fluctuations, we introduce a variable \(\sigma\), which specifies the instantaneous state of the force field. 
The force is therefore a function of both \(x\) and \(\sigma\), and for every value of \(\sigma\) it remains periodic in space,
\[
F(x + L,\sigma)  =  F(x,\sigma).
\]
\(F(x,\sigma)\) therefore represents a family of functions with the same period \(L\).

Some representative choices are
\be
F(x,\sigma) = \left\{
\begin{aligned}
&\sigma F_p(x)+F_d,\\
&F_p(x)+\sigma F_d,\\
&\sigma\left[F_p(x)+F_d\right],\\
&F_p(x)+F_d+\sigma F_{\rm swim}.
\end{aligned}
\right.
\label{eq:F0}
\ee
Here, \(F_d\) and \(F_{\rm swim}\) denote spatially uniform components, corresponding to the driving force and, 
in the case of active particles, the swimming force, respectively. The periodic component \(F_p(x)\), on the other hand, satisfies
\[
F_p(x+L)=F_p(x),
\qquad
\int_0^L dx\,F_p(x)=0,
\]
so that although \(F_p(x)\) varies within a unit cell, its average over the cell vanishes.

The first three expressions in Eq.~(\ref{eq:F0}) describe, respectively, a fluctuating periodic potential 
in the presence of a constant drive, a fixed periodic potential with a fluctuating drive, and a system 
in which both contributions fluctuate together. As a particular case, setting \(F_d=0\) and taking 
\(\sigma=\pm1\) in the first expression recovers the dichotomously switching periodic force field studied 
for a triangular potential in Refs.~\cite{PRE-Dubkov-2005,EPJB-Dubkov-2006}. The last expression describes 
a driven active particle in a periodic potential. In this case, the fluctuations are intrinsic to the active motion. 
In particular, choosing \(\sigma=\pm1\) represents one-dimensional run-and-tumble motion, with \(\sigma\) indicating 
the swimming direction~\cite{EPL-Schehr-2020,PRE-Roberts-2023}. These examples are illustrative and do not exhaust 
the possible forms of \(F(x,\sigma)\).

To completely describe the fluctuations, we still need to specify how \(\sigma\) evolves in time. In this paper, we focus on Poisson 
redraw dynamics, in which \(\sigma\) is sampled independently from \(p(\sigma)\). 
The redraws occur at a constant rate \(\tau^{-1}\), which corresponds to an exponential distribution  
\[
p_t(t_w)=\tau^{-1}e^{-t_w/\tau},
\qquad
\tau=\langle t_w\rangle,
\]
where $t_w$ is the waiting time between redraws.  
Because an exponential distribution is memoryless, the resulting dynamics is Markovian in the extended 
state space \((x,\sigma)\)~\cite{PRE-Frydel-2026}.

For a continuous distribution \(p(\sigma)\), each redraw produces a different value of \(\sigma\). For discrete states, 
the same \(\sigma\) may be drawn again. In such a case, \(\tau\) represents the mean time between redraws, not 
the mean time between actual state changes. For example, for the two-state system considered later, in which each 
state is drawn with probability \(1/2\), the mean time between state changes is \(2\tau\), rather than \(\tau\).


The evolution of the force field described above is only one possibility. Its fluctuations could instead be driven by Gaussian 
white noise~\cite{EPJB-Dubkov-2006}, or by colored noise with a finite correlation time. In this work, however, 
we limit our analysis to the Poisson redraw process described above.

\subsection*{Forward and backward Kolmogorov operators}

Although the particle is not confined and is free to move along the entire real line, \(x\in(-\infty,\infty)\), it is still possible 
to define a stationary state by taking advantage of the system’s periodicity. The real line can be partitioned into unit 
cells of size \(L\), so that the points \(x\), \(x+L\), \(x+2L\), and so on represent equivalent positions in different cells.

By accumulating the statistics from all these equivalent positions, we can fold the infinite system onto a single cell, \(x\in(0,L)\), 
with periodic boundary conditions. In this folded description, a particle leaving through \(x=L\) reenters through \(x=0\), 
and vice versa. Over time, the probability distribution within the cell approaches a stationary form.

We will subsequently characterize the periodic system in terms of this single folded cell, subject to periodic boundary conditions. 
Given the state space \((x,\sigma)\), the stationary distribution over that space is denoted by
\[
\rho\equiv\rho(x,\sigma), 
\qquad x\in(0,L), 
\]
which is normalized over a single cell as
\(
\int_0^L dx\int d\sigma\, \rho(x,\sigma) = 1,
\)
and satisfies the periodic boundary condition
\(
\rho(L,\sigma)=\rho(0,\sigma).
\)

For the subsequent analysis, we find it more convenient to work with the conditional distribution \(n(x,\sigma)\) rather than \(\rho(x,\sigma)\). 
The two distributions are related by
\be
\rho(x,\sigma) = p(\sigma)n(x,\sigma), 
\label{eq:rho-factor}
\ee
where, as mentioned earlier, \(p(\sigma)\) is the distribution of \(\sigma\). For each value of \(\sigma\), the conditional distribution is normalized as
\[
\int_0^L dx\,n(x,\sigma)=1,
\]
and, like $\rho(x,\sigma)$, obeys the periodic boundary condition $n(0,\sigma) = n(L,\sigma)$.   

\(\rho(x,\sigma)\) obeys the stationary Fokker--Planck equation, also referred to as 
the forward Kolmogorov equation. Using the relation in Eq. (\ref{eq:rho-factor}), it can 
be written in terms of the conditional distribution as
\be
0 = -\partial_x j(x,\sigma) - \frac{1}{\tau}\left[ n(x,\sigma) - \int d\sigma'\,p(\sigma')n(x,\sigma') \right],
\label{eq:Lf-fluctuating-0}
\ee
where
\be
j(x,\sigma) = \mu F(x,\sigma)n(x,\sigma) - D\partial_x n(x,\sigma)
\label{eq:j}
\ee
is the local probability flux, \(D\) is the thermal diffusion coefficient, and \(\mu\) is the particle mobility. 
The second term in Eq.~(\ref{eq:Lf-fluctuating-0}) accounts for the Poisson redraw dynamics of the force field.
Averaging Eq.~(\ref{eq:Lf-fluctuating-0}) over \(\sigma\) gives
\be
j_{\rm ss} = \int d\sigma\,p(\sigma)j(x,\sigma) = \mathrm{const}.  
\label{eq:jss}
\ee
Here, \(j_{\rm ss}\) is the stationary flux through the periodic system.

Writing the forward equation above as \(0=\hat L_f n\), we identify the forward operator as
\begin{eqnarray}
\hat L_f n 
&=& -\partial_x[\mu F  n ] + D\partial_x^2 n \nonumber\\
&& -\frac1\tau \left[ n - \int d\sigma'\,p(\sigma')\,n(x,\sigma') \right].  
\label{eq:Lf-fluctuating}
\end{eqnarray}
The corresponding backward, or adjoint, operator, acting on an arbitrary function \(g(x,\sigma)\), is given by
\begin{eqnarray}
\hat L_b g 
&=&   \mu F\, \partial_x g  + D\partial_x^2 g \nonumber\\
&& -  \frac1\tau \left[ g - \int d\sigma'\,p(\sigma')\,g(x,\sigma') \right].   
\label{eq:Lb-fluctuating}
\end{eqnarray}

We next draw attention to the similarities between the two operators.
The diffusion and Poisson redraw terms are identical; only the drift term changes under the adjoint operation.
The invariance of the redraw term follows from the reversibility of the internal-state dynamics adopted in this 
paper with respect to \(p(\sigma)\).
For more general internal dynamics, however, the internal-state term may also change when passing from 
the forward to the backward operator. In that case, reversing the force field alone would not be enough 
to construct the dual system.

Note that by taking the derivative of the backward operator, \(\partial_x[\hat L_b g]\), 
we get 
\begin{eqnarray}
\partial_x [\hat L_b g] 
&=&   \partial_x [\mu F\, \partial_x g]  + D\partial_x^2 [\partial_x g] \nonumber\\
&& -  \frac1\tau \left[ \partial_x g - \int d\sigma'\,p(\sigma')\, \partial_x g(x,\sigma') \right], 
\label{eq:dLb-fluctuating}
\end{eqnarray}
which looks like the forward operator acting on \(\partial_xg\), but for a system with the force reversed, \(F\to-F\).

This means that the relation \(\partial_x[\hat L_b g]=0\) implies that \(\partial_xg\) satisfies the stationary forward 
equation of a system with the reversed force \(F\to-F\). This observation indicates a dual system, 
whose connection to the backward equation will be explored later.
Motivated by this relation, we introduce the forward operator of the dual system,
\begin{eqnarray}
\hat L_f^D n_D 
&=& \partial_x[\mu F  n_D ] + D\partial_x^2 n_D \nonumber\\
&& -\frac1\tau \left[ n_D - \int d\sigma'\,p(\sigma')\,n_D(x,\sigma') \right].  
\label{eq:Lf-fluctuating-dual}
\end{eqnarray}
The conditional stationary distribution \(n_D(x,\sigma)\) of the dual system satisfies \(\hat L_f^D n_D = 0\).  
Its local flux is defined as
\be
j_D(x,\sigma)  =  -\mu F(x,\sigma)n_D(x,\sigma)  -  D\partial_x n_D(x,\sigma).  
\label{eq:jD}
\ee
This flux differs from the flux in Eq.~(\ref{eq:j}) only through the reversal of the force field.

\section{Homogenization framework}
\label{sec:homogenization}

We next analyze the long-time dynamics using a coarse-graining method known as homogenization. 
The goal is to obtain an expression for the effective diffusion coefficient of a particle moving through 
a periodic force field. This effective behavior emerges after the particle has traveled through many cells. 
The resulting diffusion coefficient generally differs from the microscopic thermal diffusion coefficient.

The term homogenization reflects the idea of smoothing out spatial variations and short-time fluctuations, 
leaving an effective description of the particle's motion on larger scales.
Homogenization can be formulated either analytically, through a two-scale expansion of the probability density
~\cite{book-Bensoussan-1978,book-Pavliotis-2008,PD-Vergassola-1997},
or probabilistically, by working directly with the underlying stochastic process
~\cite{AP-Bhattacharya-1985,book-Pavliotis-2008}.
Here, we adopt the probabilistic formulation.

\subsection*{Modified process}

At the microscopic level, the particle position \(X_t\) evolves according to
\be
dX_t = \mu F(X_t,\sigma_t)\,dt+\sqrt{2D}\,dB_t,
\label{eq:dXt}
\ee
where \(B_t\) denotes standard Brownian motion, while \(\sigma_t\) specifies the instantaneous state of the 
fluctuating force field and evolves according to a Poisson redraw process.

We next introduce the modified process, 
\be
M_t = X_t - v_{\rm eff}t + u(X_t,\sigma_t),
\label{eq:Mt-fluctuating}
\ee
where
\be
v_{\rm eff} = Lj_{\rm ss}
\label{eq:veff}
\ee
is the effective drift velocity. It may be generated, for example, by the driving force \(F_d\) in Eq.~(\ref{eq:F0}), 
but generally differs from \(\mu F_d\) because the periodic component of the force modifies the long-time velocity. 
The function \(u(x,\sigma)\) is a periodic corrector whose role is to remove the periodic modulation that remains 
in \(X_t-v_{\rm eff}t\).


Taking the differential of Eq.~(\ref{eq:Mt-fluctuating}) yields
\be
dM_t = dX_t-v_{\rm eff}dt + du(X_t,\sigma_t).
\label{eq:dMt-sigma}
\ee
Applying It\^o's formula for jump--diffusion processes, the differential \(du(X_t,\sigma_t)\) can be written as
\[
\begin{aligned}
du(X_t,\sigma_t)
&= [\mu F(X_t,\sigma_t)\partial_xu(X_t,\sigma_t) + D\partial_x^2u(X_t,\sigma_t)]\,dt \\
&+ \sqrt{2D}\,\partial_xu(X_t,\sigma_t) \,dB_t \\
&+ [u(X_t,\sigma_t')-u(X_t,\sigma_t)]\,dN_t.
\end{aligned}
\]
The first term is the continuous deterministic contribution, the second represents the Brownian fluctuations, 
and the third term accounts for the discontinuous change in the corrector associated with the stochastic transition
at a redraw event 
\(
\sigma_t\rightarrow\sigma_t'.
\)
Here, \(N_t\) is a Poisson counting process with rate \(\tau^{-1}\) that counts the redraw 
events~\cite{book-Watanabe-1989,book-Yin-Zhu-2010}. At each event, \(\sigma_t'\) is sampled
independently from \(p(\sigma')\).

Substituting the above expression for \(du(X_t,\sigma_t)\) into Eq.~(\ref{eq:dMt-sigma}) yields
\be
\begin{aligned}
dM_t 
& = \Big[ \mu F - v_{\rm eff} + \mu F \partial_x u + D\partial_x^2u \Big]dt \\
& + \sqrt{2D}\,\bigl[1+\partial_x u(X_t,\sigma_t) \bigr]\,dB_t  \\
& + \bigl[ u(X_t,\sigma_t') - u(X_t,\sigma_t) \bigr]\,dN_t .
\end{aligned}
\label{eq:dMt-fluctuating}
\ee

\subsection*{Corrector function}

To determine precisely the corrector function \(u(x,\sigma)\), we average \(dM_t\) over the infinitesimal stochastic increments,
conditional on the current state \((X_t,\sigma_t)=(x,\sigma)\), and denote this average by
\(
\left\langle dM_t \right\rangle_{x,\sigma}.
\)
Such averaging does not modify the first term in Eq.~(\ref{eq:dMt-fluctuating}), other than setting
\((X_t,\sigma_t)=(x,\sigma)\).
On the other hand, the Brownian term vanishes since, on average, it has no systematic contribution,
\(
\langle dB_t\rangle=0.
\)

To average the last term associated with switching, we need to specify \(dN_t\) more precisely.  
If \(N_t\) is defined as a Poisson counting process with rate \(\tau^{-1}\) that counts the number of redraws up to time \(t\), 
then the increment
\(
dN_t = N_{t+dt} - N_t
\)
indicates whether a switch occurs during the interval \([t,t+dt]\).  Such an increment has two possible values,
\be
dN_t =
\begin{cases}
1, & \text{if a redraw occurs during } [t,t+dt],\\
0, & \text{otherwise}.
\end{cases}
\label{eq:dN}
\ee
Whenever \(dN_t=1\), the post-switch state \(\sigma_t'\) is sampled from the distribution \(p(\sigma')\).
For exponentially distributed waiting times with mean \(\tau\),
\[
\langle dN_t\rangle = \frac{dt}{\tau},
\qquad
(dN_t)^2 = dN_t.
\]

We also need to average the switching term, since the transition \(\sigma\to\sigma'\) may lead to many possible states
\(\sigma'\) distributed according to \(p(\sigma')\). Its conditional average is therefore
\ba
&&
\left\langle
\bigl[u(X_t,\sigma_t')-u(X_t,\sigma_t)\bigr]dN_t
\right\rangle_{x,\sigma}
\nonumber\\
&&\qquad =
\frac{dt}{\tau}
\int d\sigma'\,p(\sigma')
\bigl[u(x,\sigma')-u(x,\sigma)\bigr].
\ea
Combining the three terms, we obtain
\be
\begin{aligned}
\left\langle dM_t \right\rangle_{x,\sigma}
&=\Bigl[ \mu F - v_{\rm eff} + \mu F\partial_xu + D\partial_x^2u \Bigr]dt \\
&\quad+
\frac{dt}{\tau}
\int d\sigma'\,p(\sigma')
\bigl[u(x,\sigma')-u(x,\sigma)\bigr].
\end{aligned}
\label{eq:dMt-fluctuating-ave}
\ee
Using the definition of the backward operator given in Eq.~(\ref{eq:Lb-fluctuating}), the above expression can be written compactly as
\be
\left\langle dM_t \right\rangle_{x,\sigma}
=
\Bigl[
\mu F(x,\sigma)-v_{\rm eff}
+\hat L_bu(x,\sigma)
\Bigr]dt.
\label{eq:dMt-ave}
\ee
The quantity in brackets represents the systematic contribution to the modified process. However, when we introduced \(M_t\),
we required it to have no systematic contribution, that is,
\[
\left\langle dM_t\right\rangle_{x,\sigma}=0.
\]
Consequently, the corrector function must satisfy
\be
\hat L_bu(x,\sigma) = v_{\rm eff} - \mu F(x,\sigma).
\label{eq:poisson-fluctuating}
\ee
We thus find that the corrector function is determined by a Poisson equation involving the backward operator.

%

\subsection*{Local diffusion}

As the auxiliary process \(M_t\) contains no systematic contributions, its most interesting aspect is its fluctuations.
To characterize these fluctuations, we begin with an expression for \((dM_t)^2\).
Squaring Eq.~(\ref{eq:dMt-fluctuating}) and retaining terms up to order \(dt\) yields
\ba
(dM_t)^2 &=& 2D[1+\partial_xu(X_t,\sigma_t)]^2dt
\nonumber\\
&+& \bigl[u(X_t,\sigma_t')-u(X_t,\sigma_t)\bigr]^2dN_t .
\label{eq:dMt2-fluctuating}
\ea
Averaging this expression conditional on the current state \((x,\sigma)\) yields the mean quadratic increment,
\ba
\frac{\langle (dM_t)^2\rangle_{x,\sigma}}{dt} &=& 2D[1+\partial_xu(x,\sigma)]^2 \nonumber\\
&+& \frac{1}{\tau}\int d\sigma'\,p(\sigma') \bigl[u(x,\sigma')-u(x,\sigma)\bigr]^2 .
\label{eq:dMt2-average}
\ea
Writing the above result as
\[
\langle (dM_t)^2\rangle_{x,\sigma}
=2D_{\rm loc}(x,\sigma)\,dt,
\]
we identify the local diffusion coefficient
\ba
D_{\rm loc}(x,\sigma)
&=&
D[1+\partial_xu(x,\sigma)]^2
\nonumber\\
&&+
\frac{1}{2\tau}\int d\sigma'\,p(\sigma')
\bigl[u(x,\sigma')-u(x,\sigma)\bigr]^2, 
\label{eq:Dloc-fluctuating}
\ea
associated with the process \(M_t\). Eq.~(\ref{eq:Dloc-fluctuating}) is the first important result of this work.
It shows how the local fluctuation strength is distributed throughout the periodic cell. The expression contains
two distinct contributions: a continuous Brownian contribution,
\(
D(1+\partial_xu)^2,
\)
and a discrete redraw contribution,
\(
\frac{1}{2\tau}\int d\sigma'\,p(\sigma')
\bigl[u(x,\sigma')-u(x,\sigma)\bigr]^2.
\)

\section{From the corrector function to the dual system}
\label{sec:dual-system}

The local diffusion coefficient \(D_{\rm loc}\) in Eq.~(\ref{eq:Dloc-fluctuating}) is expressed in terms 
of the corrector function.  
This function obeys a differential equation involving the backward operator, but beyond this mathematical role, 
it has no immediate physical interpretation, and the expression for \(D_{\rm loc}\) remains somewhat abstract.
In this section, we use the ideas of stochastic duality to provide such an interpretation.

We begin by differentiating the corrector equation, Eq.~(\ref{eq:poisson-fluctuating}), with respect to \(x\). 
After rearranging the resulting terms, we obtain
\be
0=\partial_x\!\left[\mu F n_D\right]
+D\partial_x^2n_D
-\frac{n_D}{\tau}
+\frac{1}{\tau}\int d\sigma'\,p(\sigma')n_D(x,\sigma'), 
\label{eq:FPS-nd}
\ee
where we have introduced
\be
n_D(x,\sigma) = \frac{1+\partial_xu(x,\sigma)}{L}.
\label{eq:nD}
\ee
The right-hand side of Eq.~(\ref{eq:FPS-nd}) corresponds exactly to the dual forward operator defined in
Eq.~(\ref{eq:Lf-fluctuating-dual}). Thus, the equation can be written compactly as
\[
0=\hat L_f^D n_D.
\]
Moreover, because the corrector is periodic, \(u(L,\sigma)=u(0,\sigma)\), \(n_D\) is normalized for each value of \(\sigma\):
\[
\int_0^L dx\,n_D(x,\sigma) = 1.  
\]

Together, the stationary equation \(\hat L_f^D n_D=0\) and the normalization above identify \(n_D(x,\sigma)\) 
as the conditional stationary distribution of the dual system, whose force field is reversed,
\(
F(x,\sigma)\rightarrow-F(x,\sigma).
\)
The corrector is therefore no longer merely an auxiliary function introduced by homogenization. Eq.~(\ref{eq:nD}) 
shows that its spatial derivative determines the stationary distribution of the dual system. 
This correspondence is another key result of this work.

Using Eq.~(\ref{eq:nD}) allows us to write the first term in Eq.~(\ref{eq:Dloc-fluctuating}) as
\(
D(1+\partial_xu)^2 = DL^2n_D^2(x,\sigma).
\)
The second term in Eq.~(\ref{eq:Dloc-fluctuating}) contains differences of the corrector between different force states. 
To express these differences in terms of an observable of the dual system, we rearrange Eq.~(\ref{eq:poisson-fluctuating}) as
\be
u(x,\sigma)  =  -\tau Lj_D(x,\sigma)  -  \tau v_{\rm eff}  +  \int d\sigma'\,p(\sigma')u(x,\sigma'),
\ee
where $j_D$ is the flux of the dual system defined earlier in Eq. (\ref{eq:jD}).
Taking the difference between this relation for two states, \(\sigma\) and \(\sigma'\), 
eliminates the terms that are independent of the force state and gives
\be
u(x,\sigma) - u(x,\sigma')  =  -\tau L\Bigl[j_D(x,\sigma) - j_D(x,\sigma')\Bigr].
\label{eq:uu}
\ee
Combining this result with Eq.~(\ref{eq:nD}), we can express the local diffusion coefficient entirely in terms of stationary observables of the dual system:
\ba
D_{\rm loc}(x,\sigma)
&=& D L^2 n_D^2(x,\sigma) \nonumber\\
&&+ \frac{\tau L^2}{2} \int d\sigma'\,p(\sigma') \Bigl[ j_D(x,\sigma')-j_D(x,\sigma) \Bigr]^2 . \nonumber\\
\label{eq:Dloc-dual-current}
\ea
Equation~(\ref{eq:Dloc-dual-current}) completes the dual interpretation of the local diffusion coefficient 
by eliminating the corrector in favor of stationary quantities of the dual system. It also provides the starting 
point for deriving the effective diffusion coefficient in the next section.


\section{The effective diffusion coefficient}
\label{sec:Deff}

The local diffusion coefficient \(D_{\rm loc}(x,\sigma)\) describes the strength of fluctuations at 
a particular position and force state. We now show how these local fluctuations combine to produce 
the effective diffusion coefficient governing the long-time spreading of the particle.

The effective diffusion coefficient is formally defined as
\be
D_{\rm eff} 
= \lim_{t\to\infty} \frac{\langle X_t^2\rangle-\langle X_t\rangle^2}{2t} 
= \lim_{t\to\infty} \frac{\langle(X_t-v_{\rm eff}t)^2\rangle}{2t},
\label{eq:Deff-def}
\ee
where the second equality follows from
\[
v_{\rm eff} = \lim_{t\to\infty} \frac{\langle X_t\rangle}{t}.
\]
Since the modified process is defined as
\(
M_t = X_t - v_{\rm eff}t + u(X_t,\sigma_t),
\)
and \(u(x,\sigma)\) is periodic and remains bounded, the effective diffusion coefficient can equivalently be written as
\be
D_{\rm eff} = \lim_{t\to\infty} \frac{\langle M_t^2\rangle}{2t}.
\label{eq:Deff-M}
\ee

To calculate \(\langle M_t^2\rangle\), we use
\(
d(M_t^2) = 2M_t\,dM_t + (dM_t)^2.
\)
After integration over time and averaging, the first term on the right-hand side vanishes because \(dM_t\) has zero conditional mean,
\(
\langle dM_t\rangle_{x,\sigma}=0.
\)
Then using Eq.~(\ref{eq:dMt2-fluctuating}), we obtain
\ba
\langle M_t^2\rangle &=& 2D \left\langle \int_0^t ds\, [1+\partial_xu(X_s,\sigma_s)]^2 \right\rangle \nonumber\\
&+& \left\langle \int_0^t \frac{ds}{\tau} \int d\sigma' \, p(\sigma') \bigl[ u(X_s,\sigma') - u(X_s,\sigma_s) \bigr]^2 \right\rangle .  \nonumber
\label{eq:M2-expanded}
\ea
Using the definition of the local diffusion coefficient in Eq.~(\ref{eq:Dloc-fluctuating}), 
the above result simplifies to 
\be
\langle M_t^2\rangle = 2\left\langle \int_0^t ds\,D_{\rm loc}(X_s,\sigma_s) \right\rangle .
\label{eq:M2-Dloc}
\ee

Under the standard assumption that the stochastic process \((X_t,\sigma_t)\) is ergodic, 
the long-time average of the local diffusion coefficient is equal to its stationary average,
\[
\lim_{t\to\infty} \frac1t \left\langle \int_0^t ds\,D_{\rm loc}(X_s,\sigma_s) \right\rangle = \int_0^L dx \int d\sigma\, \rho(x,\sigma) D_{\rm loc}(x,\sigma).
\]
Since 
\( 
D_{\rm eff} = \lim_{t\to\infty}\frac{\langle M_t^2\rangle}{2t}, 
\) 
this relation, together with Eq.~(\ref{eq:M2-Dloc}), yields
\be
D_{\rm eff} = \int_0^L dx  \int d\sigma\, \rho(x,\sigma) D_{\rm loc}(x,\sigma).
\label{eq:Deff-fluctuating}
\ee
Thus, the macroscopic effective diffusion coefficient emerges from the accumulation of local fluctuations, 
weighted by the stationary distribution of the system.

Substituting Eq.~(\ref{eq:Dloc-dual-current}) into Eq.~(\ref{eq:Deff-fluctuating}), the effective diffusion coefficient becomes
\ba
D_{\rm eff} &=& D L^2 \int_0^L dx \int d\sigma\, p(\sigma) n(x,\sigma) n_D^2(x,\sigma) \nonumber\\
&& + \frac{\tau L^2}{2} \int_0^L dx \int d\sigma\, p(\sigma) n(x,\sigma) \nonumber\\
&&\times \int d\sigma'\, p(\sigma') \Bigl[ j_D(x,\sigma') - j_D(x,\sigma) \Bigr]^2 .
\label{eq:Dloc-dual-current-2}
\ea
Eq.~(\ref{eq:Dloc-dual-current-2}) completes the dual reformulation of the homogenization result. 
The corrector function no longer appears explicitly; instead, \(D_{\rm eff}\) is expressed 
entirely in terms of stationary quantities of the physical and dual systems.

The significance of this expression is not merely that it provides a new representation of \(D_{\rm eff}\). 
More fundamentally, it replaces the auxiliary corrector of homogenization with physically meaningful 
stationary distributions and currents through stochastic duality.

\section{The auxiliary function \(Q(x)\) and an alternative representation of \(D_{\rm eff}\)}
\label{sec:Q}

In this section, we derive an alternative representation of \(D_{\rm eff}\) from 
Eq.~(\ref{eq:Dloc-dual-current-2}). For this purpose, we introduce the auxiliary function
\begin{align}
Q(x)
&=
D\int d\sigma\,p(\sigma)\,n(x,\sigma)n_D(x,\sigma)
\nonumber\\
&-
\tau\int d\sigma\,p(\sigma)\,j(x,\sigma)j_D(x,\sigma).
\label{eq:Q-continuous}
\end{align}
The definition of \(Q(x)\) is symmetric with respect to the physical and dual systems, 
\(
Q(x;F)=Q(x;-F).
\)
As shown in Appendix~\ref{app:dQ}, \(Q(x)\) satisfies the local balance relation
\begin{equation}
\partial_x Q(x)
=
j_{\rm ss}\Bigl[\bar n(x)-\bar n_D(x)\Bigr],
\label{eq:dQ}
\end{equation}
where
\[
\bar n(x)=\int d\sigma\,p(\sigma)n(x,\sigma),
\qquad
\bar n_D(x)=\int d\sigma\,p(\sigma)n_D(x,\sigma)
\]
are the marginal stationary densities of the physical and dual systems, respectively.

In general, \(Q(x)\) is spatially varying. Of particular interest is the case \(j_{\rm ss}=0\), 
for which Eq.~(\ref{eq:dQ}) implies
\[
\partial_x Q(x)=0,
\qquad
Q(x)=\mathrm{const}.
\]
Thus \(Q(x)\) becomes spatially uniform whenever the stationary flux vanishes.

In terms of the auxiliary function \(Q(x)\), the effective diffusion coefficient can now be written as
\ba
D_{\rm eff}[F]
&=&
L^2\int_0^L dx\,\bar n_D(x)Q(x)
\nonumber\\
&+&
L^2\tau j_{\rm ss}
\int_0^L dx\int d\sigma\,p(\sigma)\,
n(x,\sigma)j_D(x,\sigma).
\label{eq:Deff-3A}
\ea
The derivation is presented in Appendix~\ref{app:Deff3}.

\section{The case \(j_{\rm ss}=0\): generalized Lifson--Jackson relation}
\label{sec:LJ}

The representation in Eq.~(\ref{eq:Deff-3A}) is particularly useful when \(j_{\rm ss}=0\). 
Under this condition, the second term vanishes, while Eq.~(\ref{eq:dQ}) implies that \(Q(x)=Q\) is independent 
of position. The first term can therefore be written as
\(L^2Q\int_0^L dx\,\bar n_D(x)\), and since \(\int_0^L dx\,\bar n_D(x)=1\), we obtain
\be
D_{\rm eff}=L^2Q.
\ee

Using the definition of \(Q\), this result can equivalently be written as
\be
D_{\rm eff} = L^2\int d\sigma\,p(\sigma) \Bigl[ D\,n(x,\sigma)n_D(x,\sigma) - \tau j(x,\sigma)j_D(x,\sigma) \Bigr].
\label{eq:generalized-LJ}
\ee
For a non-fluctuating periodic potential, the condition \(j_{\rm ss}=0\) implies equilibrium, and the corresponding 
result reduces to the classical Lifson--Jackson relation, derived in 1962~\cite{JCP-Lifson-1962}. 
Eq.~(\ref{eq:generalized-LJ}) extends this zero-current relation to fluctuating periodic 
force fields. We therefore refer to it as the generalized Lifson--Jackson relation.
Eq.~(\ref{eq:generalized-LJ}) is one of the central results of this work.

One useful feature of Eq.~(\ref{eq:generalized-LJ}) is that its right-hand side is independent of position and can therefore be 
evaluated at any \(x\) within the unit cell. We will later exploit this property to simplify the calculations.

For later use, we write Eq.~(\ref{eq:generalized-LJ}) for a system in which \(\sigma\) takes discrete values,
\(\sigma\in\{\sigma_1,\sigma_2,\ldots\}\), with probabilities \(p_k\) satisfying \(\sum_kp_k=1\). 
The integral over \(\sigma\) is then replaced by a sum:
\be
D_{\rm eff} = L^2\sum_kp_k \Bigl[ D\,n_k(x)n_k^D(x) - \tau j_k(x)j_k^D(x) \Bigr],
\label{eq:generalized-LJ-discrete}
\ee
where
\(
n_k(x)=n(x,\sigma_k),
\)
\(
n_k^D(x)=n_D(x,\sigma_k),
\)
and similarly for the probability currents.

\section{Symmetry relations between the original and dual systems}
\label{sec:symmetry}

The dual representation raises a natural question: how do the transport coefficients 
transform under force reversal, \(F\rightarrow-F\)? Using intuition alone, we expect the effective drift velocity 
to change sign and the effective diffusion coefficient to remain unchanged,
\[
v_{\rm eff}[-F] = -v_{\rm eff}[F],
\qquad
D_{\rm eff}[-F]=D_{\rm eff}[F].
\]
However, these relations are not immediately apparent from Eq.~(\ref{eq:Dloc-dual-current-2}) or Eq. (\ref{eq:Deff-3A}).  
In this section, we establish those relations explicitly.

\subsection{Establishing \(v_{\rm eff}[-F]=-v_{\rm eff}[F]\)}

To establish the first symmetry, we begin by rewriting the corrector equation, Eq.~(\ref{eq:poisson-fluctuating}), 
in the form
\[
-Lj_D(x,\sigma) - \frac{u(x,\sigma)}{\tau} + \frac{1}{\tau}\int d\sigma'\,p(\sigma')u(x,\sigma') = Lj_{\rm ss}[F],
\]
where we used the identities \(1+\partial_xu = Ln_D\), \(\partial_x^2u = L\partial_xn_D\), the relation \(v_{\rm eff}[F]=Lj_{\rm ss}[F]\), 
and the definition of the dual current in Eq.~(\ref{eq:jD}).

Multiplying the above equation by \(p(\sigma)\) and integrating over the force states, the two terms involving \(u\) cancel. 
Dividing the resulting equation by \(L\) gives
\be
j_{\rm ss}[F] = -\int d\sigma\,p(\sigma)\,j_D(x,\sigma).
\label{eq:v-eff-dual-average}
\ee
The integral on the right-hand side is the stationary probability current of the dual system,
\be
j_{\rm ss}[-F] = \int d\sigma\,p(\sigma)\,j_D(x,\sigma).
\label{eq:v-eff-dual-dual}
\ee
Comparing these two relations gives 
\(
j_{\rm ss}[-F] = -j_{\rm ss}[F].
\)
Since \(v_{\rm eff}=Lj_{\rm ss}\), it follows that 
\[
v_{\rm eff}[-F] = -v_{\rm eff}[F].
\]

\subsection{Establishing \(D_{\rm eff}[-F]=D_{\rm eff}[F]\)}

%


To establish the second symmetry, we use Eq. (\ref{eq:Deff-3A})
Under force reversal $F\to -F$, Eq. (\ref{eq:Deff-3A}) becomes 
\ba
D_{\rm eff}[-F]
&=& L^2\int_0^L dx\,\bar n(x)Q(x) \nonumber\\
&-& L^2\tau j_{\rm ss} \int_0^L dx \int d\sigma\,p(\sigma)\, n_D(x,\sigma)j(x,\sigma),
\label{eq:Deff-3B}
\ea
where we used $j_{\rm ss} \to -j_{\rm ss}$.  
To demonstrate $D_{\rm eff}[F] = D_{\rm eff}[-F]$, we will compare each term in Eq. (\ref{eq:Deff-3A}) and Eq. (\ref{eq:Deff-3B}) 
separately.

We begin by comparing the first terms in Eqs.~(\ref{eq:Deff-3A}) and (\ref{eq:Deff-3B}). 
Their equality requires
\[
\int_0^L dx\,
\bigl[\bar n(x)-\bar n_D(x)\bigr]Q(x)=0.
\]
For \(j_{\rm ss}=0\), \(Q(x)\) is constant, and the integral vanishes because both marginal densities are normalized 
over the cell. For \(j_{\rm ss}\neq0\), Eq.~(\ref{eq:dQ}) gives
\[
\int_0^L dx\,
\bigl[\bar n(x)-\bar n_D(x)\bigr]Q(x)
=
\frac{1}{2j_{\rm ss}}\int_0^L dx\,\partial_xQ^2(x),
\]
which vanishes by periodicity. Thus, the first terms in Eqs.~(\ref{eq:Deff-3A}) and (\ref{eq:Deff-3B}) are equal.

We next compare the second terms. For their equality, it is sufficient to show that
\be
\int_0^L dx\,n_D(x,\sigma)j(x,\sigma)
+
\int_0^L dx\,n(x,\sigma)j_D(x,\sigma)
=
0.
\label{eq:term2}
\ee
This relation follows from the identity
\(
n_Dj+nj_D=-D\partial_x(nn_D),
\)
derived in Appendix~\ref{app:dQ}. Integrating it over \(x\), the right-hand side vanishes by periodicity, yielding 
Eq.~(\ref{eq:term2}).

Having demonstrated that the corresponding terms in Eqs.~(\ref{eq:Deff-3A}) and (\ref{eq:Deff-3B}) are equal, we conclude that
\[
D_{\rm eff}[F]=D_{\rm eff}[-F].
\]
Thus, the effective diffusion coefficient can be calculated using either of the two expressions.

\section{Non-fluctuating periodic force field}
\label{sec:no-switching}

While the first part of this work develops a framework for calculating \(D_{\rm eff}\) in fluctuating periodic force fields, 
the remainder applies it to several specific examples. We begin with the classical case of a non-fluctuating periodic force field.


When a constant driving force is present, \(D_{\rm eff}\) for this scenario can be expressed in terms of first-passage statistics. 
Let \(t_w\) denote the time required for the particle to travel from \(x\) to \(x+L\). Then~\cite{PRL-Reimann-2001,PRE-Reimann-2002}
\be
D_{\rm eff} = \frac{L^2}{2} \frac{\langle t_w^2\rangle-\langle t_w\rangle^2} {\langle t_w\rangle^3}.
\label{eq:Deff-fpt}
\ee
In a coarse-grained description, \(t_w\) can be viewed as the waiting time before the particle makes a directed transition 
to the neighboring cell. Since the force field is static and periodic, the periods of time a particle spends in a cell before 
moving to the neighboring cell in the direction of the driving force are independent and identically distributed.

Eq.~(\ref{eq:Deff-fpt}) follows from a coarse-grained description in which each step corresponds to a first passage by one period 
in the direction of the driving force. An alternative coarse-grained description, allowing transitions to either neighboring cell, 
has also been developed~\cite{Lindner-FNL-2001,PRE-Akimoto-2025}. Although the two descriptions involve different renewal 
processes, they lead to the same expression for the effective diffusion coefficient.

We now rederive these results using the duality framework developed in this work. 
For a non-fluctuating force field, there is only one force state, and the redraw contribution vanishes. 
Eq.~(\ref{eq:Dloc-dual-current-2}) then reduces to
\be
D_{\rm eff}=DL^2\int_0^L dx\,n(x)n_D^2(x).  
\label{eq:Deff-static}
\ee
For a static tilted periodic potential, the stationary densities \(n(x)\) and \(n_D(x)\) can be obtained explicitly. Substituting them 
into Eq.~(\ref{eq:Deff-static}) reproduces the exact effective diffusion coefficient derived in Refs.~\cite{PRL-Reimann-2001,PRE-Reimann-2002}, 
providing a consistency check of the present framework.

In the absence of the driving force, \(j_{\rm ss}=0\).  In this case, we can use the generalized
Lifson--Jackson formula in Eq.~(\ref{eq:generalized-LJ}), which yields 
\be
D_{\rm eff}=DL^2n(x)n_D(x).
\label{eq:LJ0}
\ee
With neither driving nor force-field fluctuations, the system is in equilibrium. The stationary densities of the physical and dual systems are therefore
\[
n(x)=\frac{e^{-\beta\Phi_p(x)}}{Z},
\qquad
n_D(x)=\frac{e^{\beta\Phi_p(x)}}{Z_D},
\]
where
\(
Z=\int_0^L dx\,e^{-\beta\Phi_p(x)}, 
\)
\(
Z_D=\int_0^L dx\,e^{\beta\Phi_p(x)},
\)
and $\Phi_p(x)$ is the periodic potential.  Consequently, Eq.~(\ref{eq:LJ0}) becomes
\be
D_{\rm eff}=\frac{DL^2}{ZZ_D},
\label{eq:LJ}
\ee
which recovers the classical Lifson--Jackson expression~\cite{JCP-Lifson-1962}.

\section{Two-state model \(\sigma\in\{-1,1\}\)}
\label{sec:2-state}

In the next two sections, we consider two models that satisfy \(j_{\rm ss}=0\). 
This allows us to use the generalized Lifson--Jackson relation, for which, as we recall, the auxiliary function \(Q(x)\) 
becomes position independent. One way to realize the condition \(j_{\rm ss}=0\) is through the family of systems
\[
F(x,\sigma)=\sigma F(x),
\qquad
p(\sigma)=p(-\sigma).
\]
In this case, \(j_{\rm ss}=0\) follows from the symmetry of the fluctuations, regardless of the spatial symmetry of \(F(x)\).
Indeed, replacing \(F(x,\sigma)=\sigma F(x)\) by \(F(x,\sigma)=-\sigma F(x)\) produces exactly the same system, 
by virtue of the symmetry of the fluctuations, $p(\sigma)=p(-\sigma)$. Consequently,
\(
j_{\rm ss}[-F]=j_{\rm ss}[F].
\)
And since we previously established that
\[
j_{\rm ss}[-F]=-j_{\rm ss}[F],
\]
the two relations can hold simultaneously only if \(j_{\rm ss}=0\).

In this section, we consider a simple realization of such a system: a two-state model with
\[
F(x,\sigma)=\sigma F(x),
\qquad
\sigma\in\{-1,1\},
\]
so that the force field takes one of the two forms \(F(x)\) and \(-F(x)\). The two states are drawn with equal probability,
\(
p_+=p_-=1/2.
\)

A special feature of the model \(\sigma\in\{-1,1\}\) is that the dual system is also a physical realization of the same model. 
If the stationary densities of the two states are denoted by \(n_+(x)\) and \(n_-(x)\), their dual densities satisfy
\[
n_+^D(x)=n_-(x),
\qquad
n_-^D(x)=n_+(x),
\]
and similarly for the currents,
\[
j_+^D(x)=j_-(x),
\qquad
j_-^D(x)=j_+(x).
\]
We therefore refer to this model as self-dual.

As discussed above, the total stationary current of this system vanishes,
\[
j_{\rm ss}=\frac{j_+(x)+j_-(x)}{2}=0.
\]
This allows us to use the generalized Lifson--Jackson relation in Eq.~(\ref{eq:generalized-LJ-discrete}), which for the present model becomes
\be
D_{\rm eff}
=
L^2\Bigl[D\,n_+(x)n_-(x)-\tau j_+(x)j_-(x)\Bigr].
\label{eq:Deff-2-state}
\ee

To evaluate $D_{\rm eff}$ from Eq.~(\ref{eq:Deff-2-state}) we require the stationary densities $n_+(x)$ and $n_-(x)$. 
These satisfy the coupled forward equations
\ba
&&
0   =   -\partial_x \!   \left( \mu F n_+ \right)   +   D\partial_x^2 n_+ - \frac{1}{2\tau}(n_+   -   n_-),  \nonumber\\
&&
0   =   +\partial_x \!  \left(\mu F n_-\right)   +   D\partial_x^2 n_-   -   \frac{1}{2\tau}(n_-  -  n_+),  
\label{eq:FP-self-dual}
\ea
subject to periodic boundary conditions. The associated currents are
\[
j_+  =   \mu Fn_+ - D\partial_xn_+,
\qquad
j_-   =  -\mu Fn_- - D\partial_xn_-.  
\]

\subsection{Triangular potential}
\label{sec:triangular}

As an analytically tractable example, we consider a symmetric triangular periodic potential with force field
\be
\mu F(x)=
\begin{cases}
+v, & 0<x<L/2,\\
-v, & L/2<x<L,
\end{cases}
\label{eq:F-triangle}
\ee
where \(v\) is a constant drift velocity.  This model has been studied previously in Refs.~\cite{PRE-Dubkov-2005,EPJB-Dubkov-2006,EPJE-Calisto-2017}
and provides a useful benchmark for the present formulation.

Eq.~(\ref{eq:Deff-2-state}) can be simplified by evaluating it at a position where \(j_+=j_-=0\), so that the second term vanishes.  
For the triangular force field, \(x=0\) is one such position. Evaluating Eq.~(\ref{eq:Deff-2-state}) at this position gives
\be
D_{\rm eff} = L^2D\,n_+(0)n_-(0).
\label{eq:Deff-2-state-A}
\ee

Because the individual currents also vanish at \(x=L/2\), the interval \(0<x<L/2\) is bounded by two zero-current points 
and can therefore be treated as a system with reflecting walls. Within this interval, the stationary densities satisfy
\begin{align}
0&=-v\partial_x n_+ + D\partial_x^2 n_+
-\frac{1}{2\tau}(n_+-n_-),\nonumber\\
0&=+v\partial_x n_- + D\partial_x^2 n_-
-\frac{1}{2\tau}(n_--n_+),
\end{align}
subject to
\(f
j_\pm(0)=j_\pm(L/2)=0
\)
and since we restrict the problem to half of the unit cell, the densities are normalized as 
\(
\int_0^{L/2}dx\,n_\pm(x)=\frac12.
\)
These are the same equations 
that describe a one-dimensional run-and-tumble particle confined between two reflecting walls. Their exact stationary 
solution is available in Refs.~\cite{PRE-Razin-2020,JSTAT-Frydel-2021,PRE-Frydel-2022}.

Using the exact run-and-tumble solution of the equations above, we obtain the stationary densities \(n_+(x)\) and \(n_-(x)\). 
Substituting their values at \(x=0\) into Eq.~(\ref{eq:Deff-2-state-A}) gives
\be
\frac{D_{\rm eff}}{D} = \frac{(1+Pe)\left(Pe+\cosh^2 a\right)} {\left(\cosh a+Pe\,\frac{\sinh a}{a}\right)^2},
\label{eq:Deff-triangle}
\ee
where the dimensionless parameters are defined as
\be
Pe = \frac{\tau v^2}{D},
\qquad
\alpha = \frac{Lv}{D},
\qquad
a = \frac{\alpha}{4}\sqrt{1+\frac{1}{Pe}} .
\label{eq:parameters}
\ee
Here, \(Pe\) denotes the Péclet number.

In the limit \(\tau\to0\), the force is redrawn so rapidly that the particle feels only the average of the two force states, which for the 
self-dual system considered here is zero. 
The motion then reduces to ordinary thermal diffusion,
\[
\frac{D_{\rm eff}}{D}=1.
\]
In the opposite limit, \(\tau\to\infty\), redraw events cease and the particle remains in one of the two force states. 
Each state corresponds to a static, untilted periodic potential, and both give the same effective diffusion coefficient. 
Eq.~(\ref{eq:Deff-triangle}) then reduces to the classical Lifson--Jackson result,
\[
\frac{D_{\rm eff}}{D} = \left[ \frac{\alpha/4}{\sinh(\alpha/4)} \right]^2 <1.
\]
Another limiting case is the zero-temperature limit, \(D\to0\), for which Eq.~(\ref{eq:Deff-triangle}) reduces to
\[
D_{\rm eff} = \frac{\tau v^2}{\left(1+4\tau v/L\right)^2}.
\]
Thus, even when thermal diffusion vanishes, fluctuations of the force field give rise to a finite effective diffusion.

\begin{figure}[t]
\centering
\begin{tabular}{cc}
\includegraphics[width=0.23\textwidth]{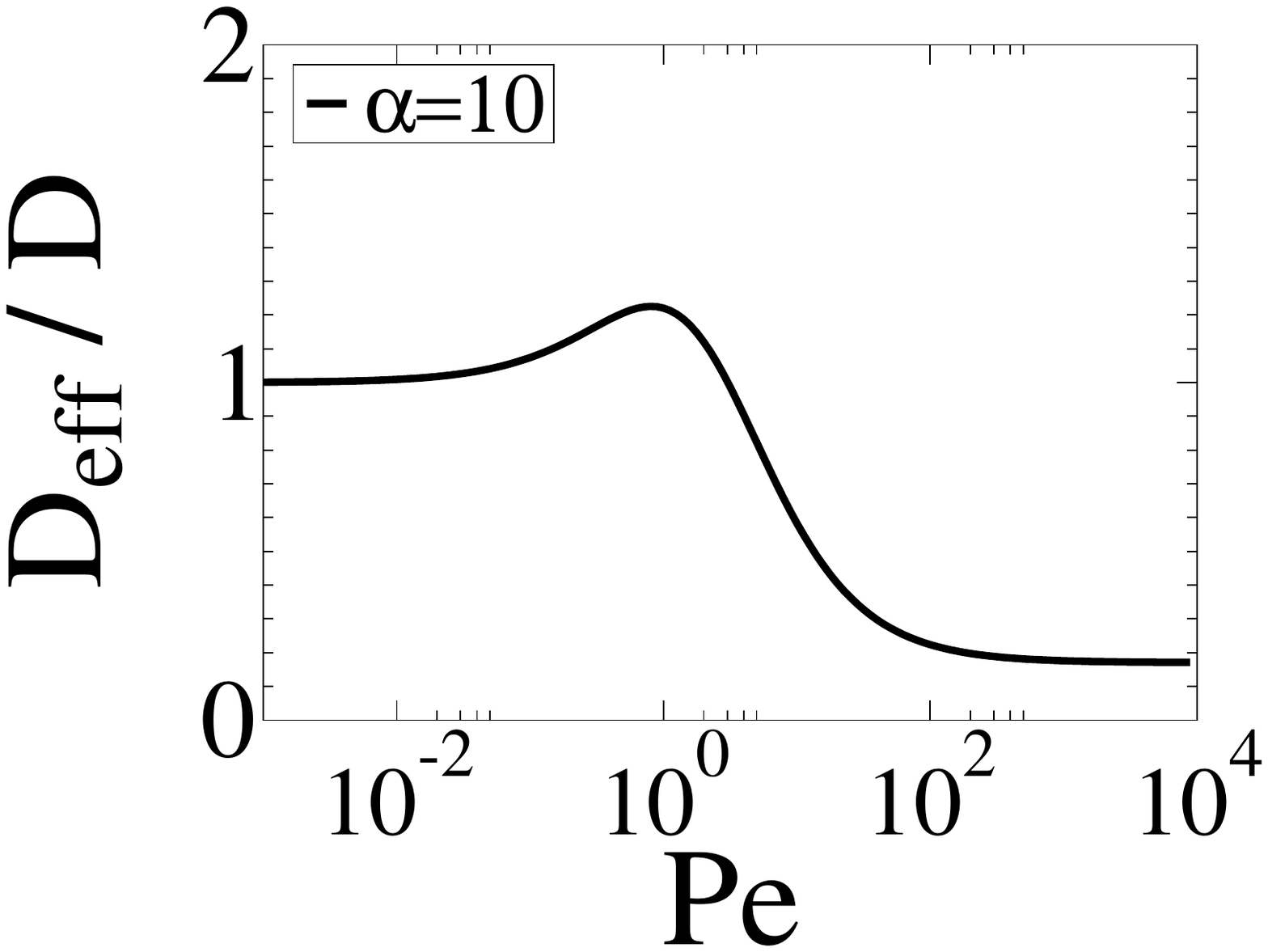} 
\\[2mm]
\includegraphics[width=0.23\textwidth]{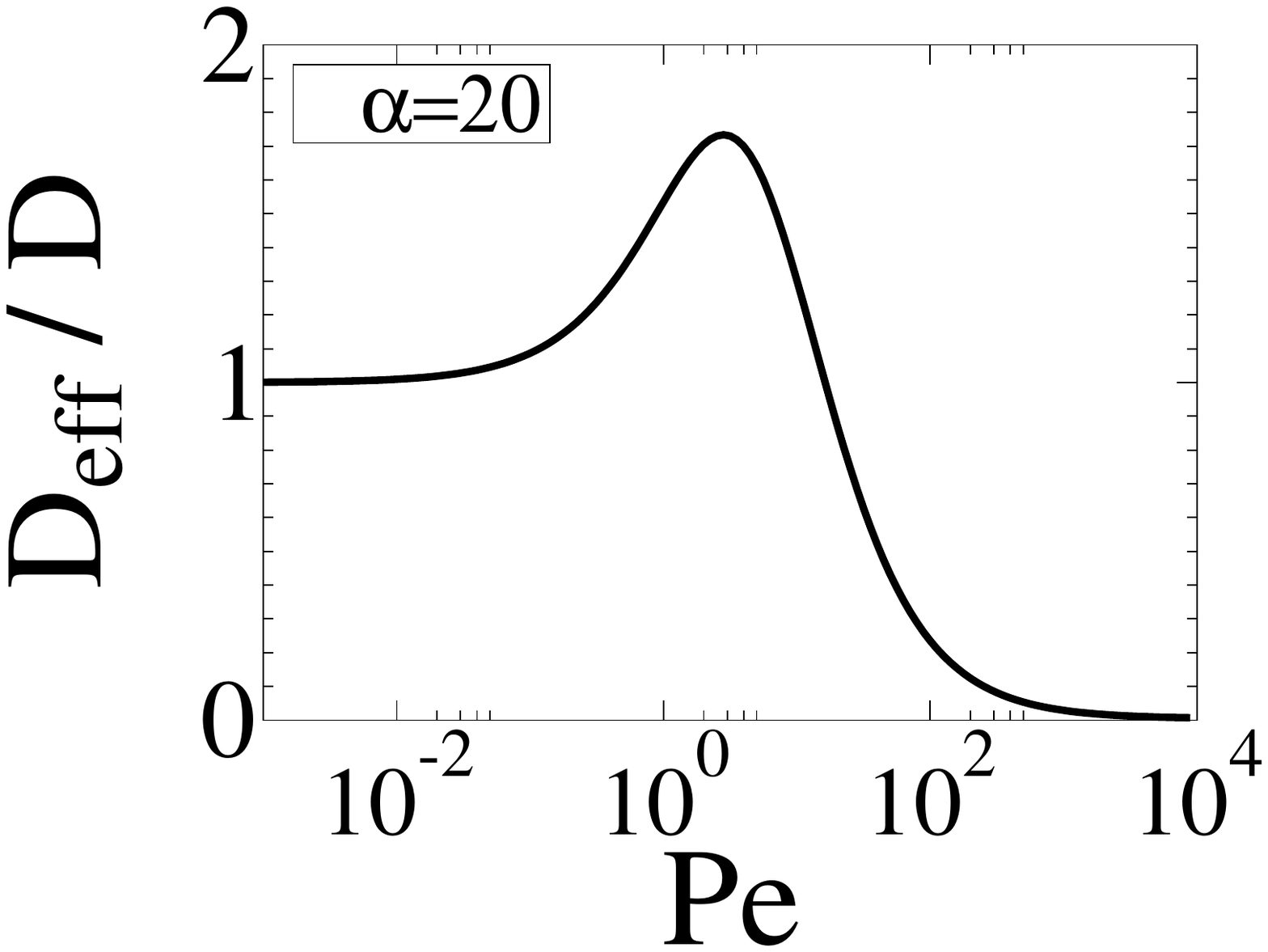} 
\\[2mm]
\includegraphics[width=0.23\textwidth]{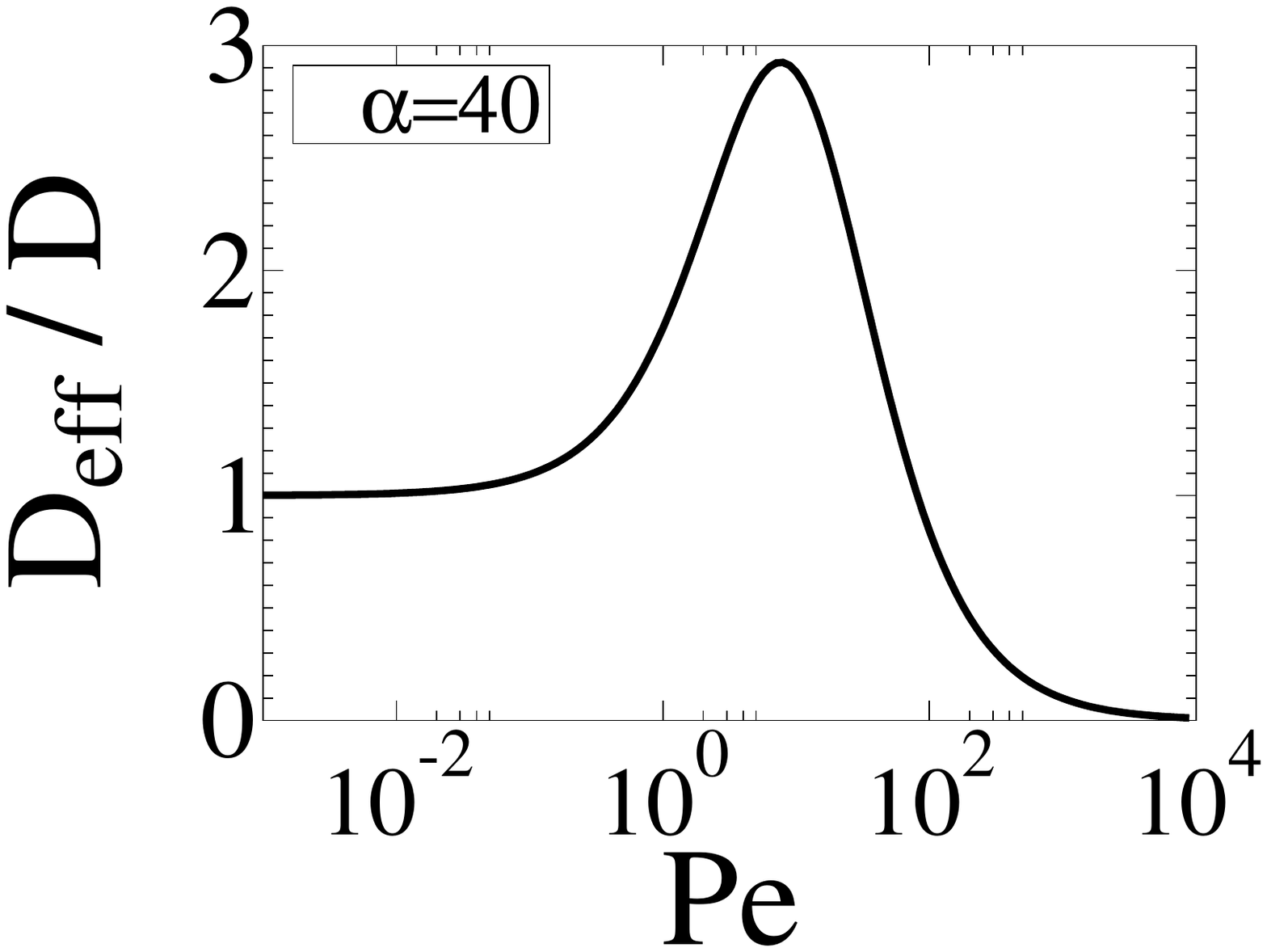} 
\end{tabular}
\caption{
The normalized effective diffusion coefficient \(D_{\rm eff}/D\) as a function of the Péclet number \(Pe\).
From top to bottom, the panels correspond to \(\alpha=10\), \(20\), and \(40\). The curves are obtained from
Eq.~(\ref{eq:Deff-triangle}).
}
\label{fig:Deff-Pe}
\end{figure}

In Fig.~\ref{fig:Deff-Pe}, we plot \(D_{\rm eff}/D\) as a function of \(Pe\) using Eq.~(\ref{eq:Deff-triangle}). 
Since \(Pe=\tau v^2/D\), varying \(Pe\) at fixed \(v\) and \(D\) is equivalent to varying the mean redraw time \(\tau\). 
All curves are nonmonotonic, indicating an optimal redraw time at which effective diffusion is maximized. 

\subsection*{Comparison with existing results}
\label{sec:comparison}

The triangular periodic potential considered above was previously analyzed in 
Refs.~\cite{PRE-Dubkov-2005,EPJB-Dubkov-2006}. The approach adopted in those works represents the periodic system as a sequence 
of identical and statistically independent cells. The particle is then described as jumping from one cell to the next, with the 
waiting time for each jump determined by the first-passage time to either neighboring cell. This independent-cell construction gives
\be
D_{\rm eff}^{\rm FP}=\frac{L^2}{2\tau_f},
\label{eq:Deff-FP-A}
\ee
where \(\tau_f\) is the mean first-passage time to reach either \(x=L\) or \(x=-L\), starting from \(x=0\).

For the fluctuating potential, the initial force state is selected as \(+\) or \(-\) with equal probability. The mean first-passage 
time used in Eq.~(\ref{eq:Deff-FP-A}) is therefore
\[
\tau_f=\frac{T_+(0)+T_-(0)}{2},
\]
where \(T_\pm(x)\) denotes the mean first-passage time to either absorbing boundary at \(x=\pm L\) for a particle initially at 
position \(x\) in state \(\pm\).

This construction corresponds to a continuous-time random walk on a lattice, in which a particle waits for a random time 
drawn from a prescribed waiting-time distribution, 
before jumping to one of the neighboring sites \cite{JMP-Montroll-1965,book-Klafter-2011}.  
The quantity \(\tau_f\) in this case is the mean waiting time of the corresponding 
distribution.  

Solving for \(T_\pm(x)\) for the triangular potential and substituting the result into Eq.~(\ref{eq:Deff-FP-A}) leads to the analytical 
expression given in Eq.~(33) of Ref.~\cite{PRE-Dubkov-2005}, or equivalently Eq.~(15) of Ref.~\cite{EPJB-Dubkov-2006}. 
That expression, however, does not agree with the exact result derived here in Eq.~(\ref{eq:Deff-triangle}).

To understand the origin of this discrepancy, we examine the assumption that successive cell-to-cell passages are statistically 
independent. If the assumption were correct, the resulting value of $D_{\rm eff}$ should be independent of the coarse-graining 
length. A coarse-grained jump may span one, two, or, more generally, $m$ spatial periods. Starting from the center of a reference 
cell at $x=0$, this corresponds to first passage to either $x=-mL$ or $x=mL$.

Accordingly, we define the first-passage formula, 
\be
D_{\rm eff}^{\rm FP}(m) = \frac{m^2L^2}{2\tau_f^{(m)}},
\label{eq:Deff-FP-m}
\ee
where \(\tau_f^{(m)}\) designates the mean first-passage time to reach either absorbing boundary at \(x=\pm mL\), 
starting from \(x=0\).  If successive cell-to-cell passages were statistically independent, this estimate would be independent 
of the coarse-graining length \(m\).

\begin{figure}[t]
\centering
\includegraphics[width=0.30\textwidth]{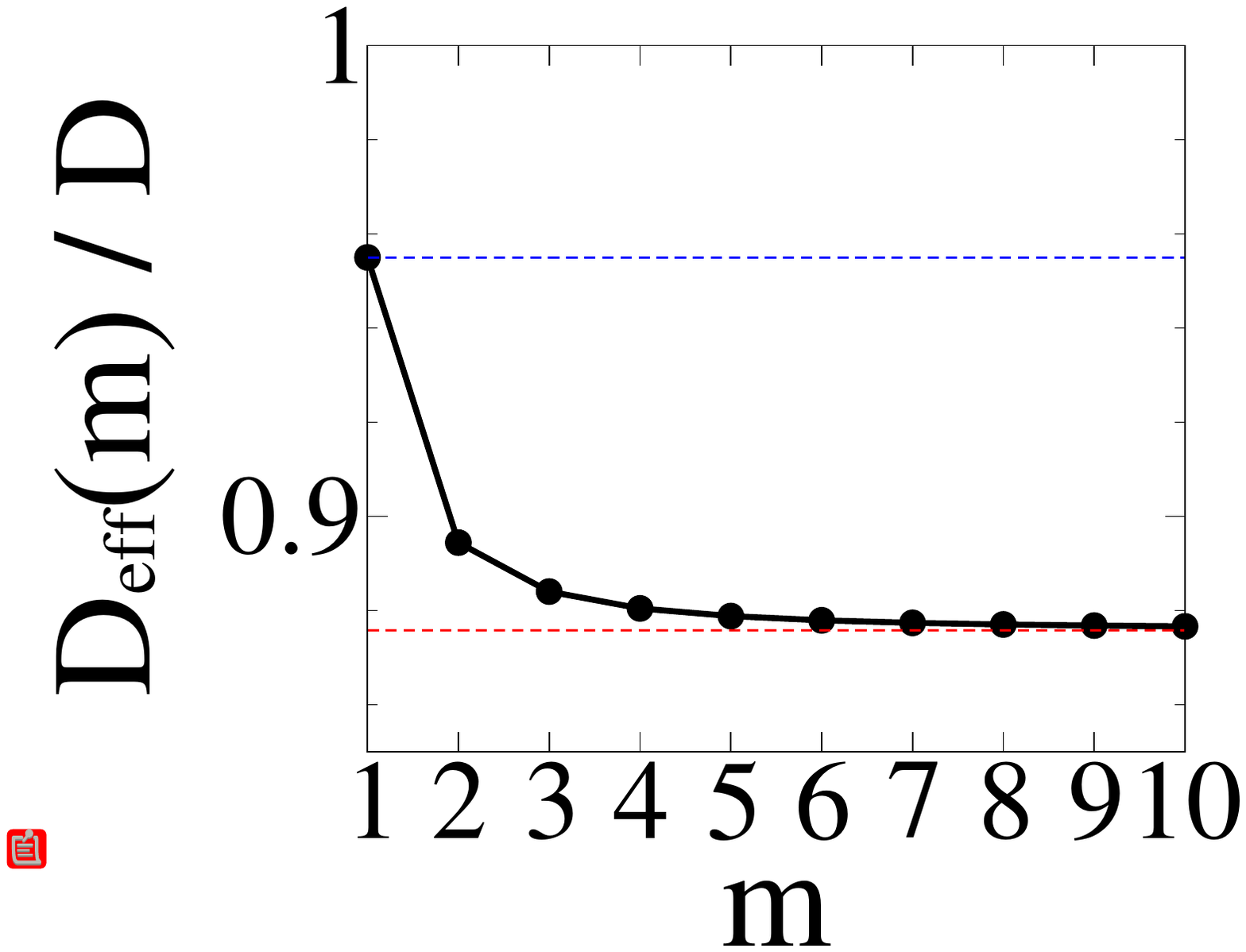}
\caption{
The first-passage estimate \(D_{\rm eff}^{\rm FP}(m)\), defined in Eq.~(\ref{eq:Deff-FP-m}), as a function
of the coarse-graining length \(m\). As \(m\) increases, \(D_{\rm eff}^{\rm FP}(m)\) converges to the exact analytical
result in Eq.~(\ref{eq:Deff-triangle}), represented by the dashed red line. The case \(m=1\) reproduces the expression reported in
Refs.~\cite{PRE-Dubkov-2005,EPJB-Dubkov-2006}, represented by the dashed blue line.
The symbols were obtained by numerically solving the coupled backward equations described in Appendix~\ref{app:MFPT-m}.
The system parameters are \(D=0.05\) and \(\tau=L=v=1\).
}
\label{fig:Deff-m}
\end{figure}

However, as Fig.~\ref{fig:Deff-m} shows, the first-passage formula depends on \(m\) and decreases according to
\[
D_{\rm eff}^{\rm FP}(1) > D_{\rm eff}^{\rm FP}(2) > D_{\rm eff}^{\rm FP}(3) > \cdots > D_{\rm eff}.  
\]
As the coarse-graining length increases, the estimate approaches the exact effective diffusion coefficient,
\be
D_{\rm eff} = \lim_{m\to\infty} D_{\rm eff}^{\rm FP}(m),
\label{eq:convergence}
\ee
given in Eq.~(\ref{eq:Deff-triangle}). The expression reported in Refs.~\cite{PRE-Dubkov-2005,EPJB-Dubkov-2006} corresponds 
to the one-cell estimate \(m=1\).

This result shows that, for the fluctuating potential considered here, the effective diffusion coefficient is not given by the 
single-cell relation
\(
D_{\rm eff}=L^2/(2\tau_f).
\)
The mean first-passage time itself is well defined; the problem lies in treating successive passages as independent lattice jumps. 
In calculating \(\tau_f\), the initial force states are assigned equal probability. At an actual cell crossing, however, 
the force state is inherited from the preceding passage, introducing correlations between successive passages. The influence of 
these correlations decreases as the coarse-graining length increases, consistent with the convergence observed in 
Fig.~\ref{fig:Deff-m}. Thus, although the independent-cell construction is suitable for a non-fluctuating periodic potential, 
it is insufficient when correlations between successive passages are present.

A complementary test of the validity of the independent-cell assumption is to shift the center of the reference cell. 
Consider a cell centered at an arbitrary position \(x_c\). Starting from \(x=x_c\), the corresponding first-passage 
problem has absorbing boundaries at
\[
x=x_c-L
\qquad\text{and}\qquad
x=x_c+L.
\]
If the independent-cell construction were valid, the resulting first-passage estimate of \(D_{\rm eff}\) should be independent of \(x_c\).

\begin{figure}[t]
\centering
\includegraphics[width=0.25\textwidth]{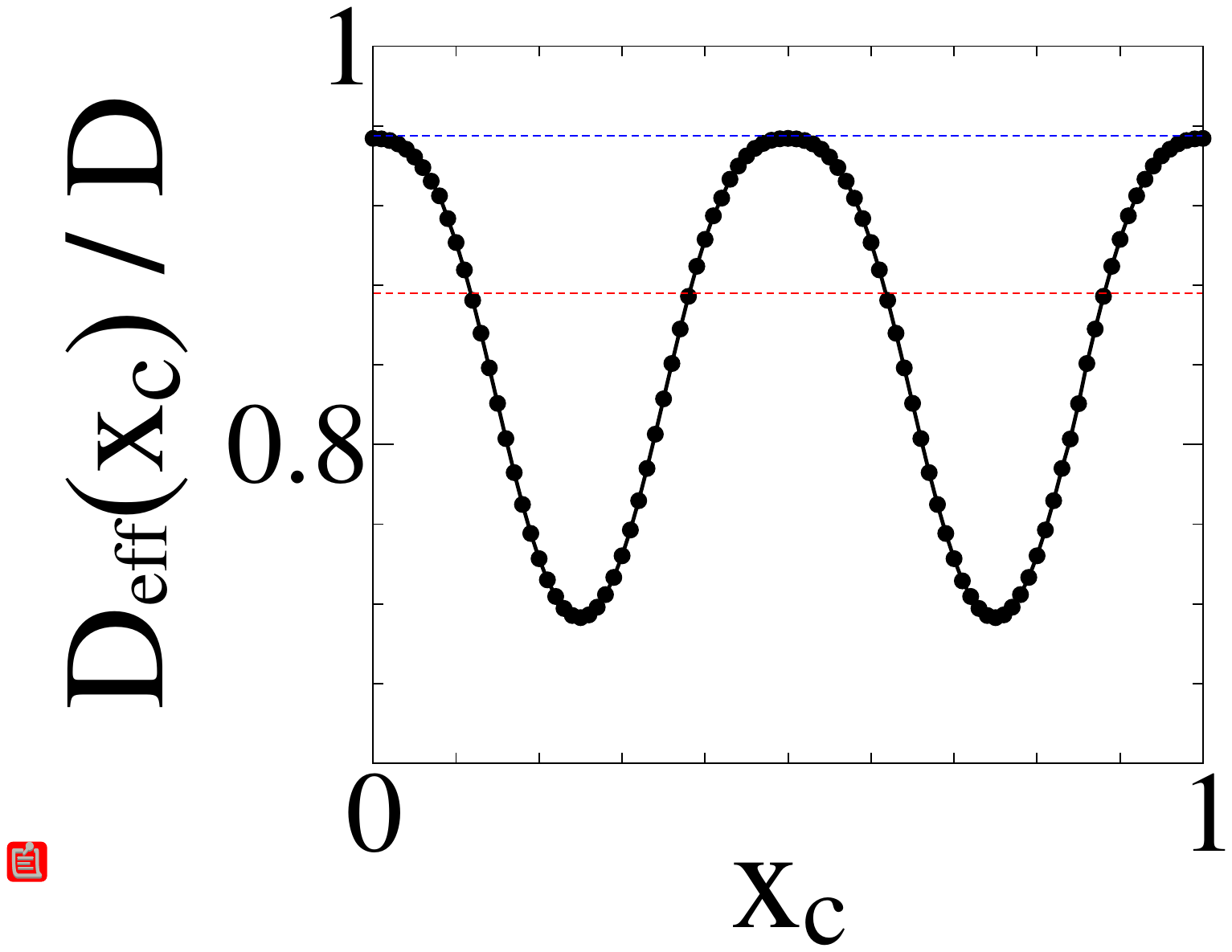}
\caption{
The single-cell first-passage estimate \(D_{\rm eff}^{\rm FP}(x_c)\) as a function of the reference-cell center \(x_c\). 
The blue and red dashed lines have the same meanings as in Fig.~\ref{fig:Deff-m}, indicating the single-cell estimate 
and the exact effective diffusion coefficient, respectively. Symbols show simulation results. The parameters are \(D=0.05\) and \(\tau=L=v=1\).
}
\label{fig:Deff-xc}
\end{figure}

Consistent with the sensitivity to coarse graining observed in Fig.~\ref{fig:Deff-m}, Fig.~\ref{fig:Deff-xc} shows that the single-cell 
estimate \(D_{\rm eff}^{\rm FP}(x_c)\) also depends on the choice of \(x_c\).

As an independent test of Eq.~(\ref{eq:Deff-triangle}), we also calculate the time-dependent diffusion coefficient directly from particle trajectories,
\be
D_{\rm eff}(t)  =  \frac{\langle X_t^2\rangle-\langle X_t\rangle^2}{2t}.
\label{eq:Deff-time}
\ee
By definition, \(D_{\rm eff}(t)\) approaches the effective diffusion coefficient in the long-time limit; see Eq.~(\ref{eq:Deff-def}). 
Its limiting value should therefore agree with the analytical result in Eq.~(\ref{eq:Deff-triangle}).

The trajectories are generated by discretizing the Langevin equation with a time step \(\Delta t\),
\[
X_{t+\Delta t}
=
X_t+\mu F(X_t,\sigma_t)\Delta t
+\sqrt{2D\Delta t}\,\xi_t,
\]
where \(\xi_t\) is an independent Gaussian random variable with zero mean and unit variance. 
The force is evaluated at the particle position \(X_t\) at the beginning of each time step. A trajectory may therefore cross a force 
discontinuity during a single update, but the resulting discretization error vanishes as \(\Delta t\to0\).

All trajectories start from \(X_0=0\), with the initial force state \(\sigma_0=\pm1\) selected with equal probability. 
At each time step, a redraw occurs with probability
\(
1-e^{-\Delta t/\tau}.
\)
When a redraw occurs, the new state is selected independently as \(\sigma=\pm1\) with equal probability.

\begin{figure}[t]
\centering
\includegraphics[width=0.48\linewidth]{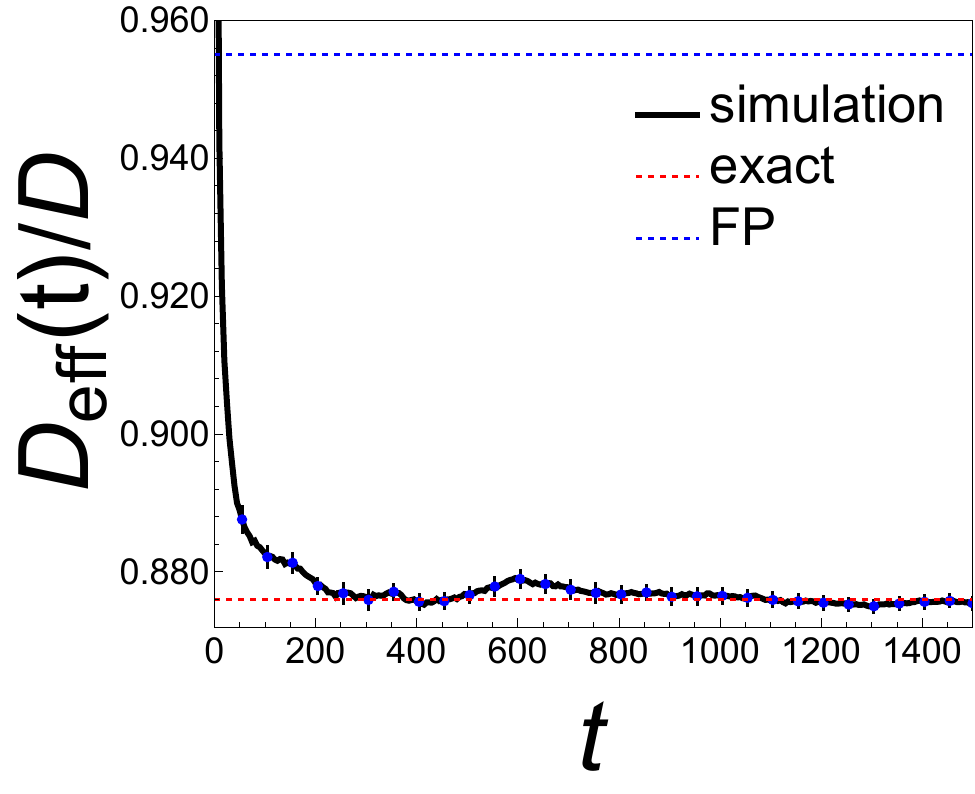}
\hspace{-0.0\linewidth}
\includegraphics[width=0.48\linewidth]{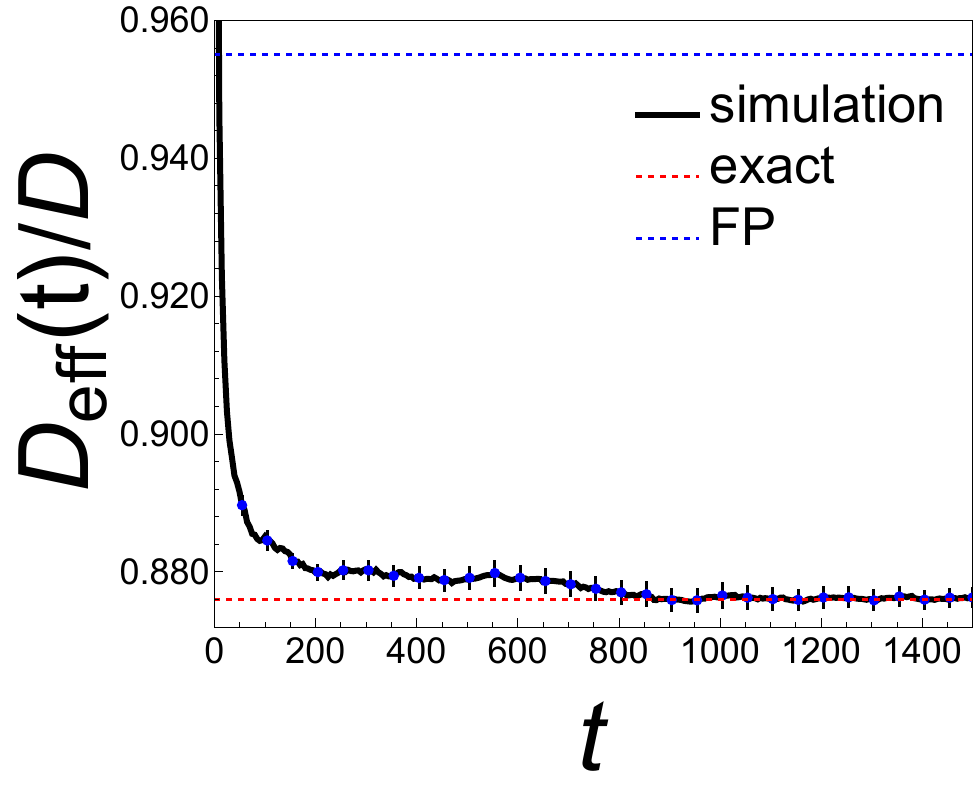}
\caption{
The time-dependent diffusion coefficient \(D_{\rm eff}(t)\), calculated directly from
\(\operatorname{Var}(X_t)/(2t)\) using \(10^6\) independent trajectories.
Panels (a) and (b) correspond to time steps \(\Delta t=10^{-3}\) and \(\Delta t=10^{-4}\), respectively.
The blue and red dashed lines have the same meanings as in Fig.~\ref{fig:Deff-m}, indicating the single-cell estimate
and the exact effective diffusion coefficient, respectively.  The horizontal axis shows the simulation time \(t\).
The parameters are \(L=v=\tau=1\) and \(D=0.05\). Error bars show the Gaussian estimate of the standard error, 
\(D_{\rm eff}(t)\sqrt{2/(N-1)}\), and do not include systematic time-discretization errors.
}
\label{fig:Deff-time}
\end{figure}

Fig.~\ref{fig:Deff-time} shows \(D_{\rm eff}(t)\) obtained from \(10^6\) independent trajectories using two time steps,
\(\Delta t=10^{-3}\) and \(\Delta t=10^{-4}\).
In both cases, the estimates converge to the analytical result, and the two simulations agree within their statistical uncertainties.
The statistical uncertainty is determined primarily by the number of trajectories 
and remains approximately constant at long times.

\section{Two-state on--off model \(\sigma\in\{0,1\}\)}
\label{sec:two-state-on-off}

In this section, we consider another model with \(j_{\rm ss}=0\). Unlike the model of the previous section, the fluctuations 
do not satisfy \(p(\sigma)=p(-\sigma)\), and the vanishing of the stationary current is therefore no longer guaranteed by the 
symmetry of the fluctuations. It can instead be ensured by an appropriate spatial symmetry of the periodic potential. A sufficient 
condition is the so-called supersymmetry condition introduced by Reimann~\cite{PRL-Reimann-2001a},
\[
\Phi(x-L/2)=-\Phi(x).
\]
The triangular potential considered in the previous section satisfies this condition, and we therefore continue to use it here.

As an example of asymmetric fluctuations, we consider on--off switching, 
\(\sigma\in\{0,1\}\). For equally probable on and off states,
\(
p_{\rm on}=p_{\rm off}=1/2,
\)
the generalized Lifson--Jackson relation in Eq.~(\ref{eq:generalized-LJ-discrete}) becomes
\be
D_{\rm eff}
=
\frac{L^2}{2}
\bigg[
D\Big(n_{\rm on}n_{\rm on}^{D}+n_{\rm off}n_{\rm off}^{D}\Big)
-
\tau\Big(j_{\rm on}j_{\rm on}^{D}+j_{\rm off}j_{\rm off}^{D}\Big)
\bigg].
\label{eq:Deff-on-off}
\ee

Since Eq.~(\ref{eq:Deff-on-off}) is independent of \(x\), we can evaluate it at any position. 
In the current model, the local currents vanish at \(x=0\) and \(x=L/2\). Thus, by evaluating Eq.~(\ref{eq:Deff-on-off}) 
at \(x=0\), we can eliminate the terms that depend on those local currents.  
Moreover, for the triangular potential, reversing the force is equivalent to shifting the system by half a period, 
which implies 
\[
n_{\rm on}^D(x)=n_{\rm on}(x+L/2),
\qquad
n_{\rm off}^D(x)=n_{\rm off}(x+L/2).
\]
Using these relations, Eq.~(\ref{eq:Deff-on-off}) evaluated at $x=0$ becomes
\be
D_{\rm eff} = \frac{DL^2}{2} \Big[ n_{\rm on}(0)n_{\rm on}(L/2) + n_{\rm off}(0)n_{\rm off}(L/2) \Big].
\label{eq:Deff-on-off-triangle}
\ee
Thus, calculating \(D_{\rm eff}\) requires only the stationary densities of the physical system at two positions, \(x=0\) and \(x=L/2\).

Since the local currents vanish at both positions, the stationary problem on \(0<x<L/2\) can be treated as a problem confined 
between two reflecting boundaries. Within this interval, the stationary densities satisfy the coupled Fokker--Planck equations
\ba
&&
0 = D\partial_x^2 n_{\rm off} - \frac{n_{\rm off}-n_{\rm on}}{2\tau}, \nonumber\\
&&
0 = -v\partial_x n_{\rm on} + D\partial_x^2 n_{\rm on} - \frac{n_{\rm on}-n_{\rm off}}{2\tau},
\ea
subject to the reflecting boundary conditions
\(
j_{\rm on}(0) = j_{\rm off}(0) = j_{\rm on}(L/2)=j_{\rm off}(L/2)=0
\)
and the normalization
\(
\int_0^{L/2}dx\,n_{\rm off}(x) = \int_0^{L/2}dx\,n_{\rm on}(x) = 1/2.
\)

The stationary densities can be expressed as linear combinations of three exponential modes \(e^{\lambda_i x/L}\), 
with dimensionless inverse lengths \(\lambda_i\) and coefficients \(c_i\). Substituting these solutions into 
Eq.~(\ref{eq:Deff-on-off-triangle}) gives
\be
\frac{D_{\rm eff}}{D}  =  \frac12 \sum_{i,j=1}^{3} c_i c_j e^{\lambda_j/2} \left[ 1 - \frac{\lambda_i+\lambda_j}{\alpha}  +  \frac{2\lambda_i\lambda_j}{\alpha^2} \right].
\label{eq:Deff-on-off-lambda}
\ee
The dimensionless inverse lengths \(\lambda_i\), \(i=1,2,3\), are the roots of
\[
Pe\,\lambda^3 - \alpha Pe\,\lambda^2 - \alpha^2\lambda + \frac{\alpha^3}{2} = 0,
\]
while the three coefficients \(c_i\) are determined by the two reflecting boundary conditions,
\[
\sum_{i=1}^{3} \lambda_i \left( 1-\frac{\lambda_i}{\alpha} \right)c_i = 0,
\qquad 
\sum_{i=1}^{3} \lambda_i \left( 1-\frac{\lambda_i}{\alpha} \right)c_i e^{\lambda_i/2} = 0,
\]
together with the normalization condition
\[
\frac{1}{\alpha} \sum_{i=1}^{3} c_i \left( e^{\lambda_i/2}-1 \right) = \frac12.
\]
The parameters \(\alpha\) and \(Pe\) were defined previously in Eq.~(\ref{eq:parameters}).

\begin{figure}[t]
\centering
\begin{tabular}{cc}
\includegraphics[width=0.23\textwidth]{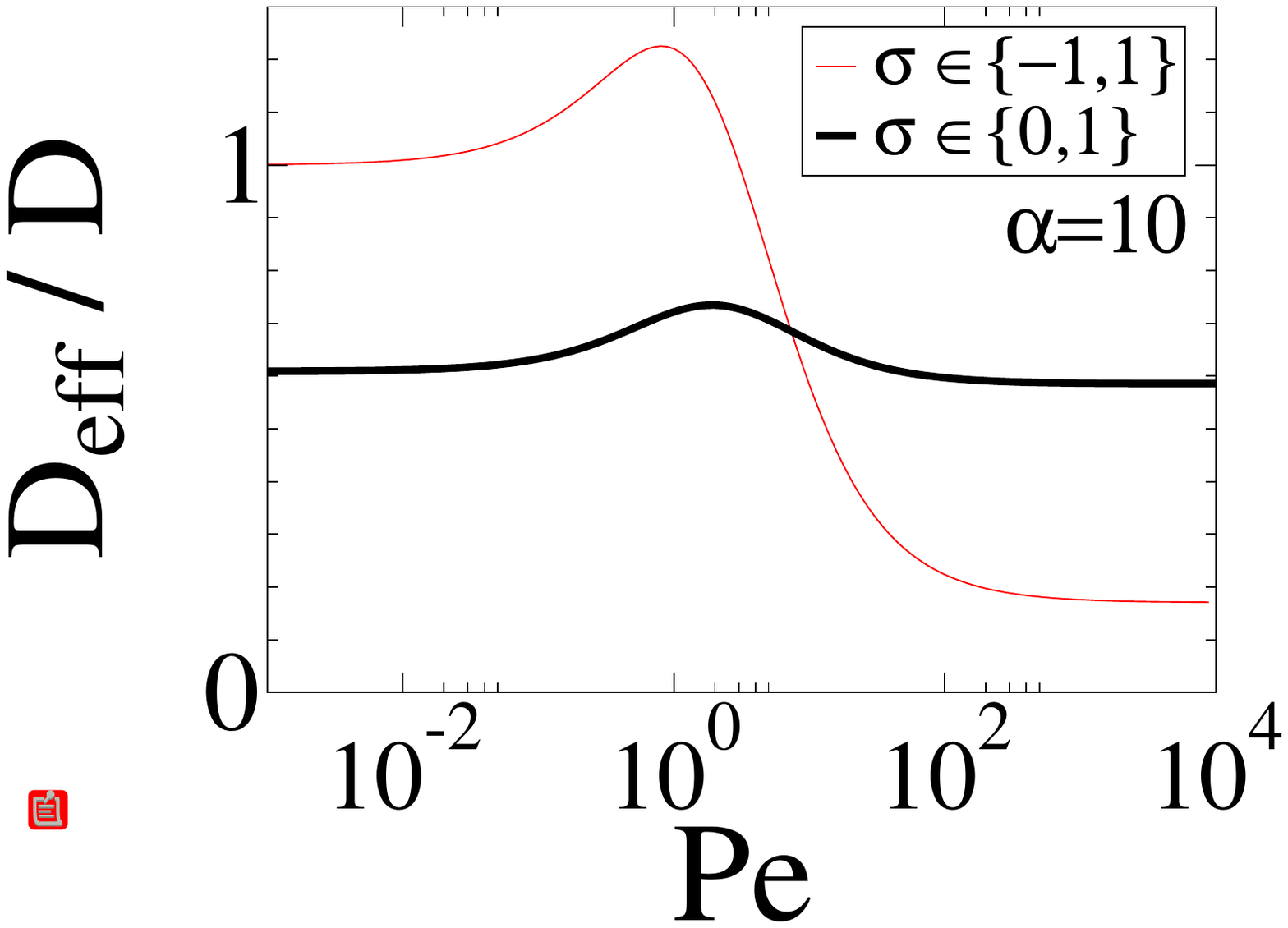} 
\\[2mm]
\includegraphics[width=0.23\textwidth]{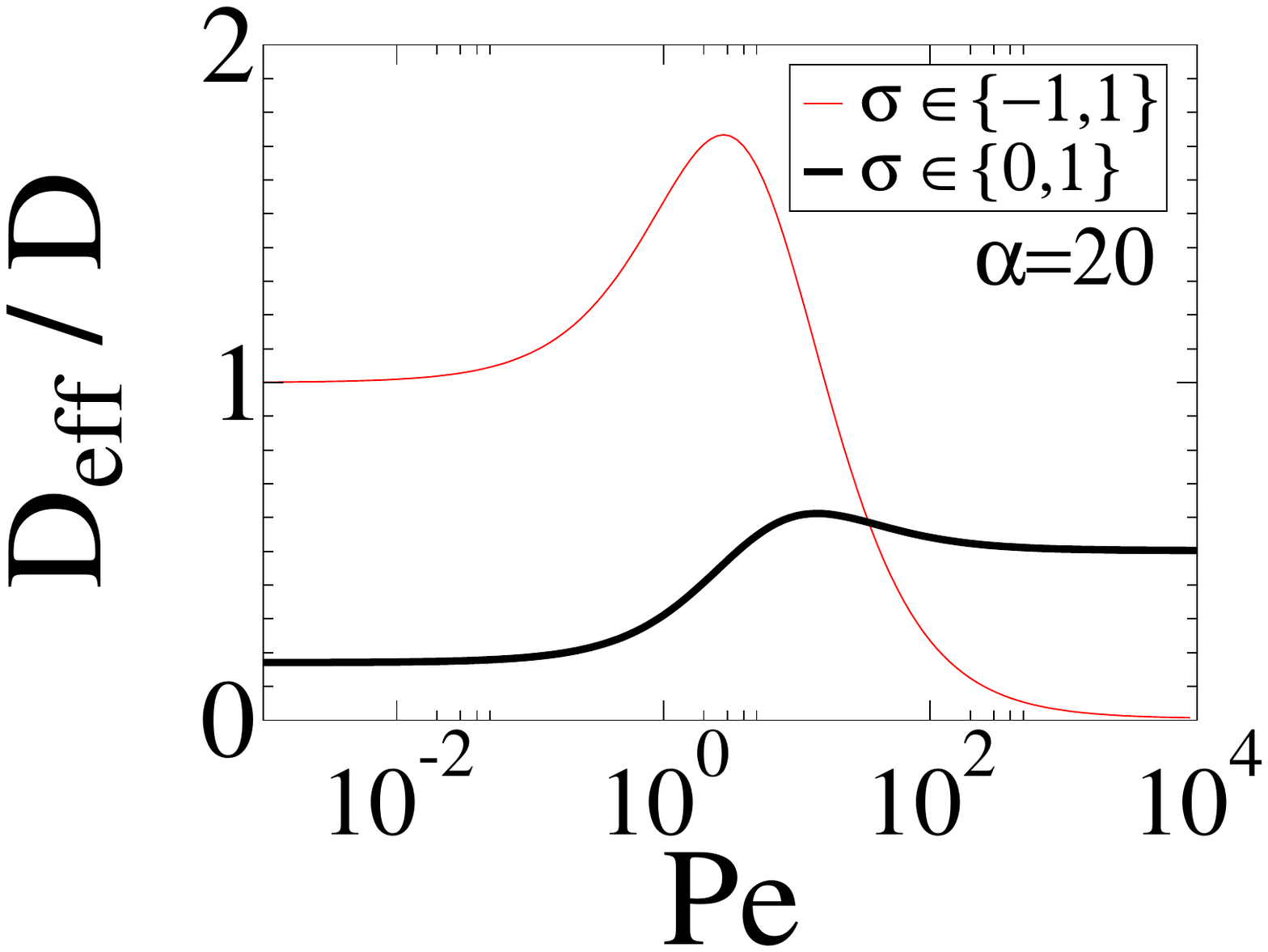} 
\\[2mm]
\includegraphics[width=0.23\textwidth]{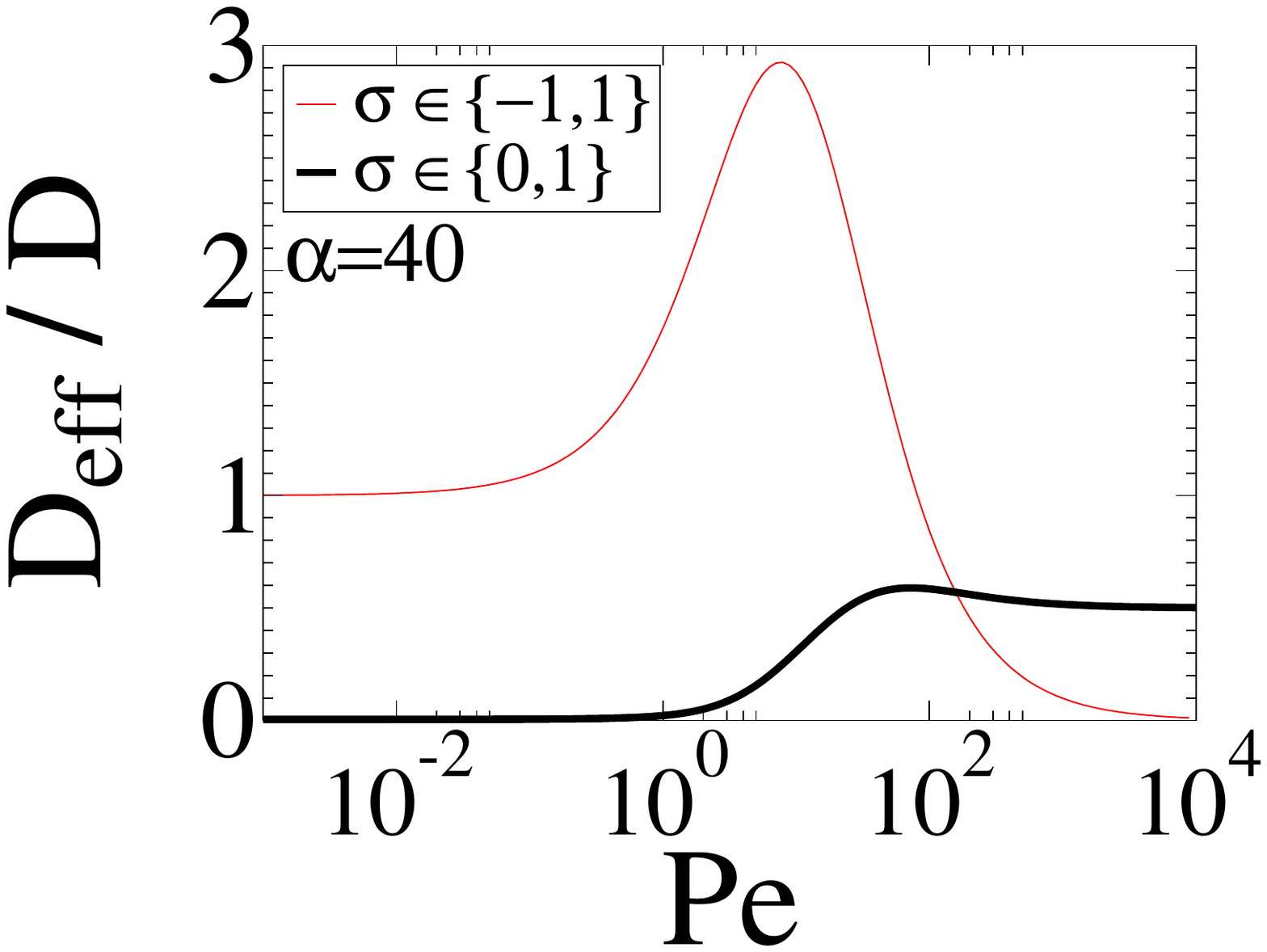} 
\end{tabular}
\caption{
The normalized effective diffusion coefficient \(D_{\rm eff}/D\) as a function of the Péclet number \(Pe\).
From top to bottom, the panels correspond to \(\alpha=10\), \(20\), and \(40\). Each panel compares the on--off model,
calculated from Eq.~(\ref{eq:Deff-on-off-lambda}), with the corresponding self-dual model described by Eq.~(\ref{eq:Deff-2-state}).
}
\label{fig:Deff-Pe-A}
\end{figure}

Fig.~\ref{fig:Deff-Pe-A} shows the resulting \(D_{\rm eff}\) as a function of \(Pe\) for several values of \(\alpha\), 
together with the corresponding results for the \(\sigma\in\{-1,1\}\) model. Unlike the \(\sigma\in\{-1,1\}\) model, 
the on--off model does not exhibit diffusion enhancement: for all values of \(Pe\) and \(\alpha\) considered here, \(D_{\rm eff}\leq D\).

\begin{figure}[t]
\centering
\includegraphics[width=0.25\textwidth]{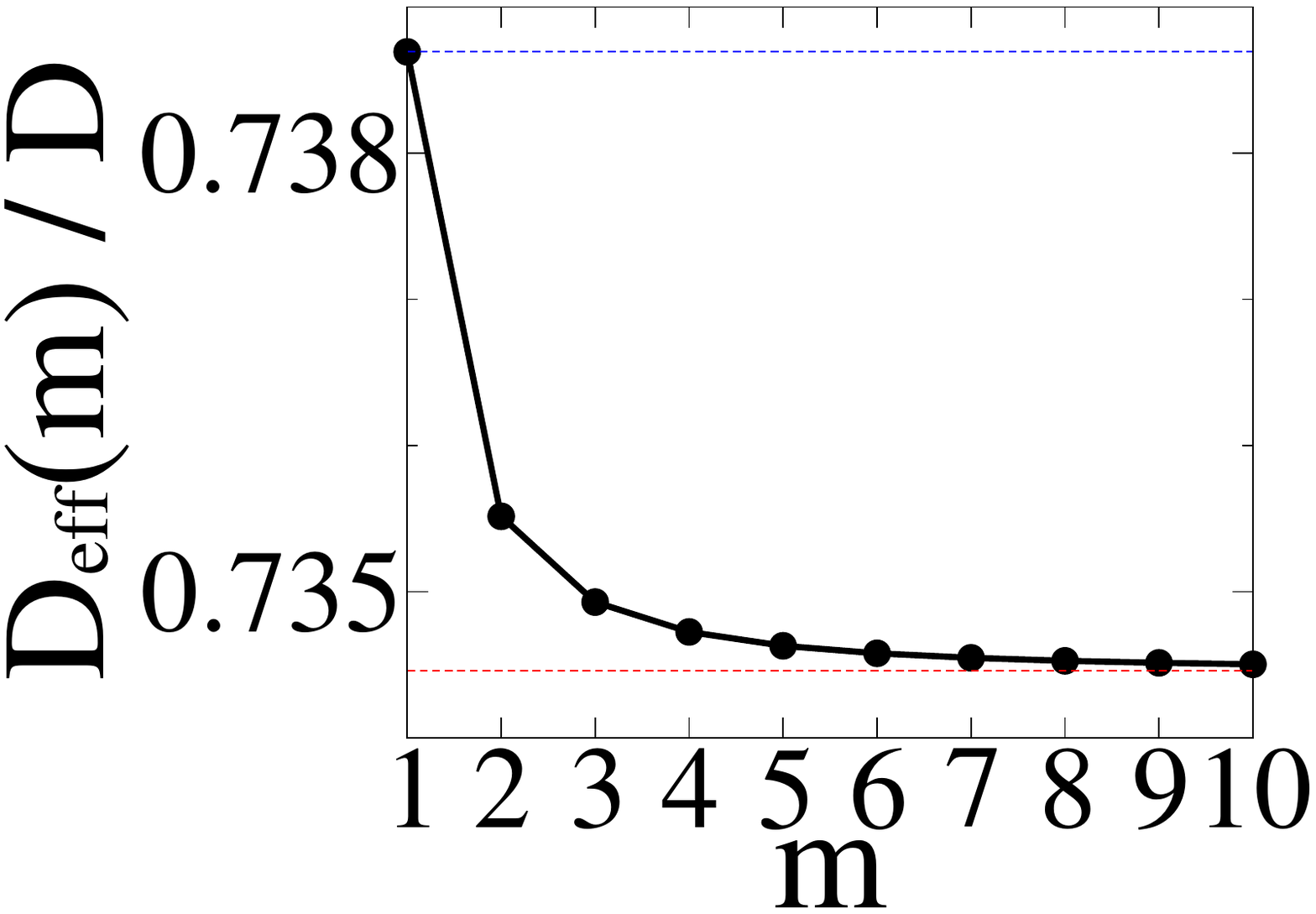}
\caption{
The normalized first-passage estimate \(D_{\rm eff}^{\rm FP}(m)/D\) as a function of \(m\), corresponding 
to a coarse-graining length \(mL\). The symbols were obtained by numerically solving the coupled backward 
equations described in Appendix~\ref{app:MFPT-m}. The dashed blue line represents the single-cell estimate 
\(D_{\rm eff}^{\rm FP}(1)/D\), whereas the dashed red line represents the analytical result in Eq.~(\ref{eq:Deff-on-off-lambda}). 
The parameters are \(D=0.1\), \(\tau=0.2\), and \(L=v=1\).
}
\label{fig:Deff-m-on-off}
\end{figure}

We next examine the first-passage estimate \(D_{\rm eff}^{\rm FP}(m)\), defined in Eq.~(\ref{eq:Deff-FP-m}), 
for different coarse-graining lengths \(mL\). The initial on and off states are assigned equal probability. 
As in the \(\sigma\in\{-1,1\}\) model, the force state at an actual cell crossing is inherited from the preceding passage, 
so the independent-cell construction neglects correlations between successive passages. The data points in 
Fig.~\ref{fig:Deff-m-on-off} show that \(D_{\rm eff}^{\rm FP}(m)\) converges gradually to the analytical result in 
Eq.~(\ref{eq:Deff-on-off-lambda}) as \(m\) increases, consistent with the analogous result in Fig.~\ref{fig:Deff-m}.

\begin{figure}[t]
\centering
\includegraphics[width=0.5\linewidth]{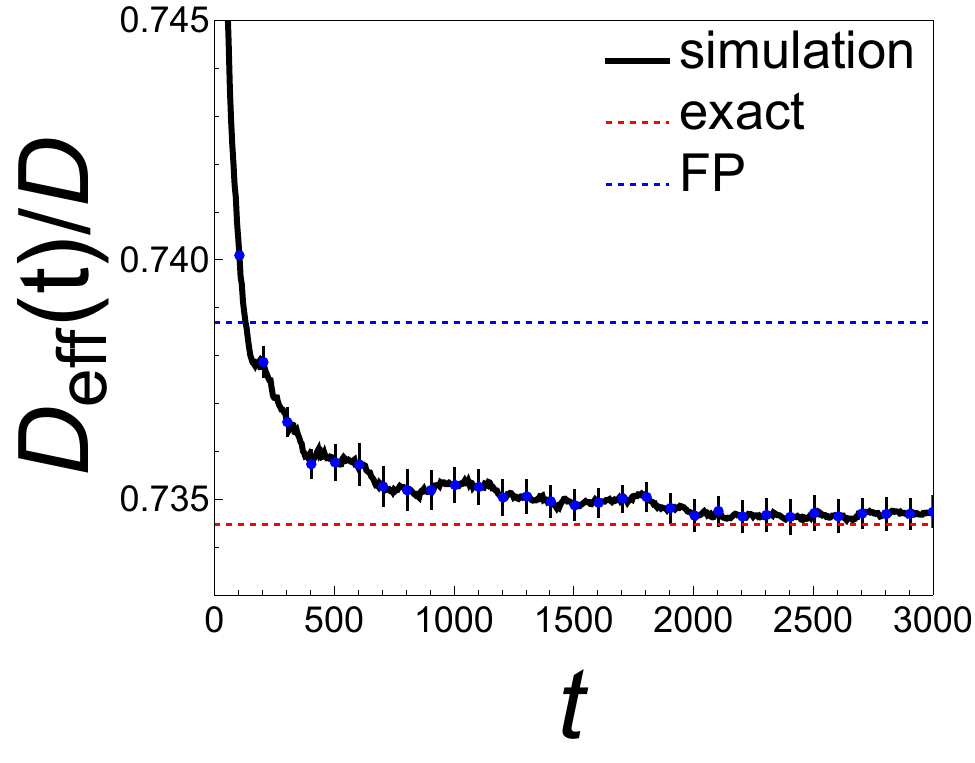}
\hspace{-0.02\linewidth}
\includegraphics[width=0.5\linewidth]{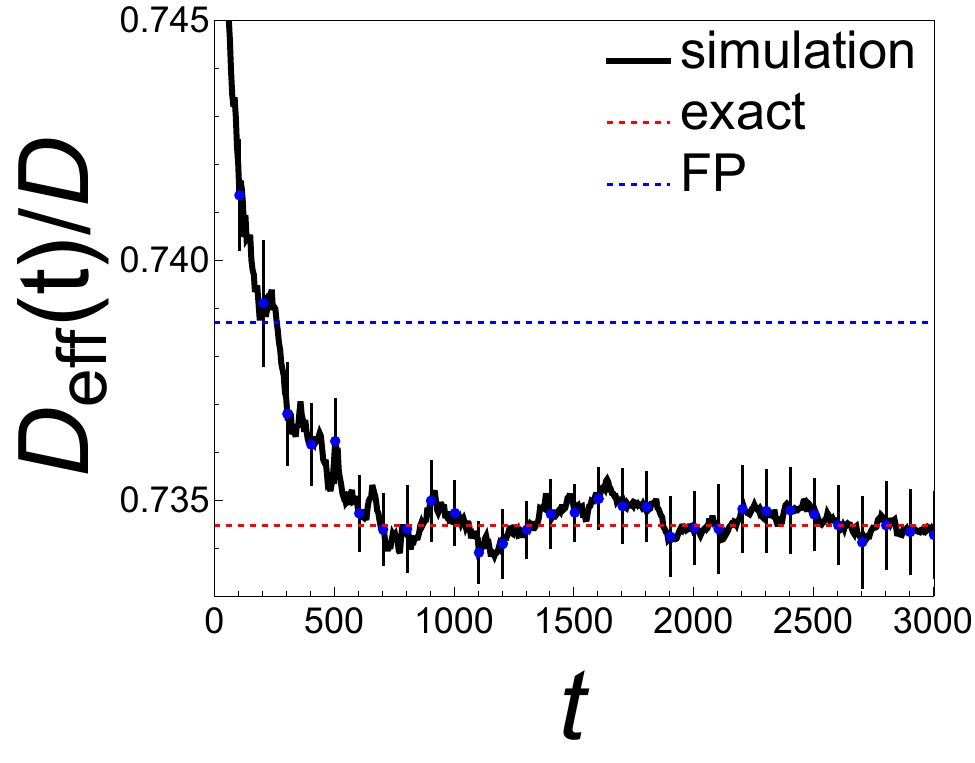}
\caption{
The normalized time-dependent diffusion coefficient \(D_{\rm eff}(t)/D\), calculated directly from particle trajectories.
Panel (a) shows results obtained using \(10^7\) independent trajectories and a time step \(\Delta t=10^{-3}\), while
panel (b) shows results obtained using \(10^6\) trajectories and a time step \(\Delta t=10^{-4}\).
The dashed blue line represents the single-cell estimate \(D_{\rm eff}^{\rm FP}(1)/D\), whereas the dashed red line
represents the analytical result in Eq.~(\ref{eq:Deff-on-off-lambda}). The horizontal axis shows the simulation time \(t\).
The parameters are \(L=v=1\), \(\tau=0.2\), and \(D=0.1\).
}
\label{fig:Deff-time-on-off}
\end{figure}

Finally, we verify Eq.~(\ref{eq:Deff-on-off-lambda}) using the time-dependent diffusion coefficient \(D_{\rm eff}(t)\), 
defined in Eq.~(\ref{eq:Deff-time}) and calculated directly from particle trajectories. Figure~\ref{fig:Deff-time-on-off} 
shows results obtained using two different time steps. In both cases, \(D_{\rm eff}(t)\) approaches the analytical result 
as \(t\) increases.

\section{Application to a finite driving force}
\label{sec:finite-driving}

The applications considered so far have primarily focused on the generalized Lifson--Jackson relation, which requires 
\(j_{\rm ss}=0\). To place those results in a more general context, we now briefly consider a driven system with 
\(j_{\rm ss}\neq0\). Specifically, we consider the two-state force field
\be
F(x,\sigma)=\sigma F_p(x)+F_d, \qquad \sigma\in\{-1,1\},
\ee
where \(F_p(x)=F_p(x+L)\) is the periodic component and \(F_d\) is a constant driving force. 
The two states occur with equal probabilities, \(p_+=p_-=1/2\). The force in the corresponding dual system is
\(
F_D(x,\sigma) = -\sigma F_p(x)-F_d.
\)

For this model, Eq.~(\ref{eq:Deff-3A}) becomes
\ba
D_{\rm eff}
&=& L^2\int_0^L dx\,\bar n_D(x)Q(x) \nonumber\\
&+& \frac{\tau L^2j_{\rm ss}}{2} \int_0^L dx\, \left[ n_+(x)j_+^D(x)+n_-(x)j_-^D(x) \right],
\label{eq:Deff-finite-driving}
\ea
where
\(
\bar n_D = ({n_+^D + n_-^D}) / {2},
\)
Because \(j_{\rm ss}\neq0\), the second term in Eq.~(\ref{eq:Deff-finite-driving}) does not vanish. 
Moreover, the auxiliary function,
\ba
Q(x) 
&=& \frac{D}{2} \left[ n_+(x)n_+^D(x) + n_-(x)n_-^D(x) \right]  \nonumber\\
&-& \frac{\tau}{2} \left[ j_+(x)j_+^D(x) + j_-(x)j_-^D(x) \right], 
\label{eq:Q-finite-driving}
\ea
is generally position-dependent and must remain inside the spatial integral.

In this case, fewer analytical simplifications are available. Nevertheless, the central result of this work 
remains applicable: \(D_{\rm eff}\) can be calculated from stationary quantities of the physical and dual systems. 
This provides an alternative to direct, computationally expensive simulations of the long-time particle displacement. 
The alternative procedure, however, requires stationary simulations of both the physical and dual systems.

\begin{figure}[t]
\centering
\includegraphics[width=0.25\textwidth]{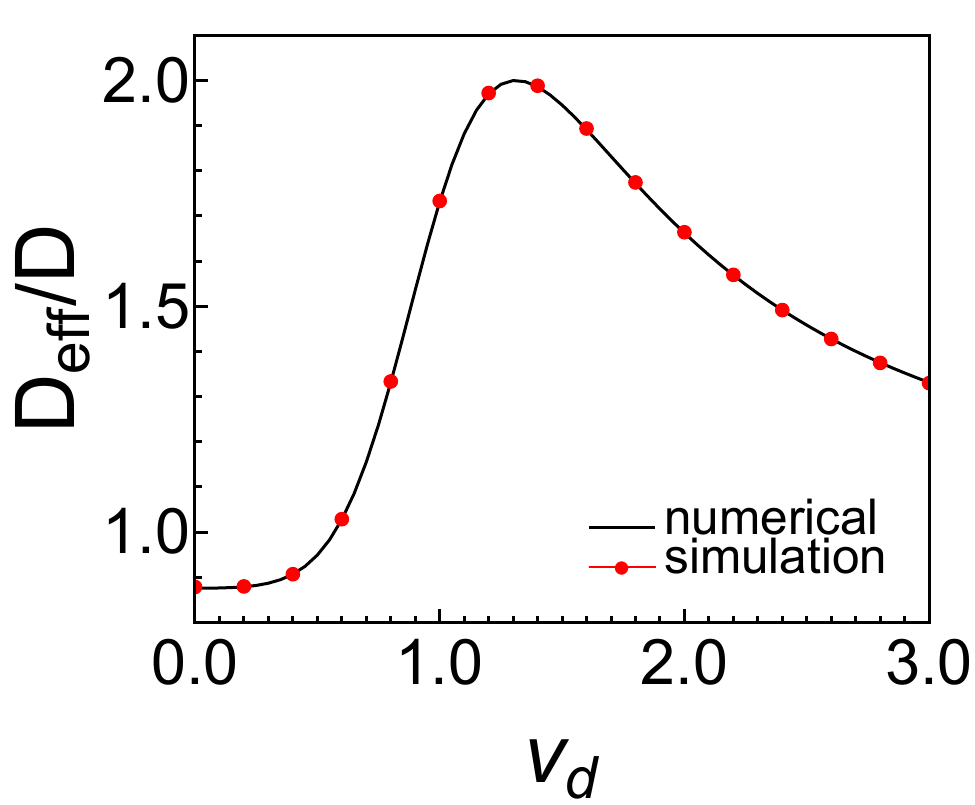}
\caption{
The normalized effective diffusion coefficient \(D_{\rm eff}/D\) as a function of the driving velocity
\(v_d=\mu F_d\). Symbols show the results obtained from Eq.~(\ref{eq:Deff-finite-driving}), using stationary
densities and currents measured in simulations of the physical and dual systems. The solid line shows the
corresponding results obtained by numerically solving the stationary equations. The simulations use
\(10^5\) particles, a time step \(\Delta t=5\times10^{-4}\), and \(400\) spatial bins per period.
The remaining parameters are \(D=0.05\) and \(\tau=L=v=1\).
}
\label{fig:Deff-vd}
\end{figure}

We calculate \(D_{\rm eff}\) from Eq.~(\ref{eq:Deff-finite-driving}) using the stationary densities of the physical and dual systems. 
The corresponding currents are reconstructed from these densities, as explained in the next paragraph. 
The stationary densities are obtained using two approaches: simulations of the physical and dual systems within a single cell 
with periodic boundary conditions, and numerical solutions of the corresponding stationary equations. 
Figure~\ref{fig:Deff-vd} compares the two results and shows that they are in good agreement. 
We point out that the stationary simulation data can be generated at a relatively low computational cost compared with the 
direct calculation of \(D_{\rm eff}\) from the long-time displacement variance.  This computational efficiency is the main practical
advantage of the alternative formulation of \(D_{\rm eff}\) in terms of the stationary densities and currents of the physical and dual systems.

In the simulations, particles were confined to a single unit cell with periodic boundary conditions. The state-resolved stationary 
densities \(n_\pm(x)\) and \(n_\pm^D(x)\) were obtained by binning particle positions within the cell. The corresponding local currents 
were reconstructed from the stationary balance equations,
\[
\partial_xj_+(x)=\frac{n_-(x)-n_+(x)}{2\tau},
\qquad
\partial_xj_-(x)=-\partial_xj_+(x).
\]
In particular,
\[
j_+(x)
=
C_+
+
\frac{1}{2\tau}
\int_0^xdy\,\bigl[n_-(y)-n_+(y)\bigr].
\]
The integration constant \(C_+\) was determined by integrating the constitutive relation
\[
j_+(x)=\mu F(x,+)n_+(x)-D\partial_xn_+(x)
\]
over one period. The integrated diffusive term vanishes because \(n_+(x)\) is periodic. The analogous procedure was used to obtain 
\(j_-(x)\), as well as the currents \(j_\pm^D(x)\) of the dual system. This method avoids taking numerical derivatives of the sampled 
density profiles.

\begin{figure}[t]
\centering
\includegraphics[width=0.5\linewidth]{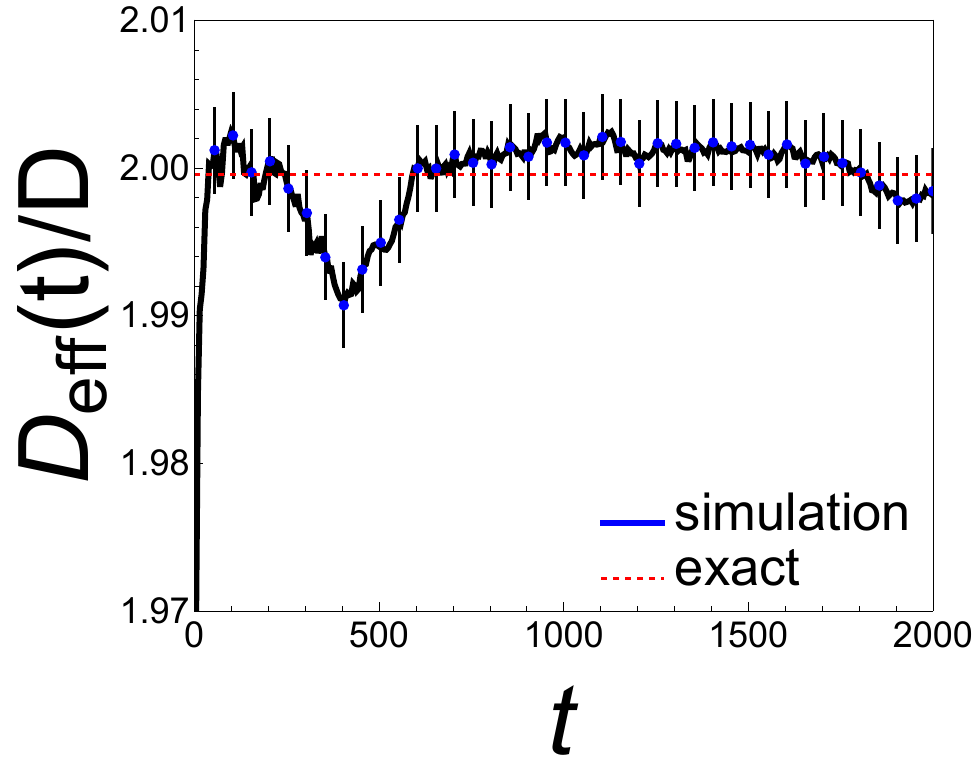}
\hspace{-0.02\linewidth}
\includegraphics[width=0.5\linewidth]{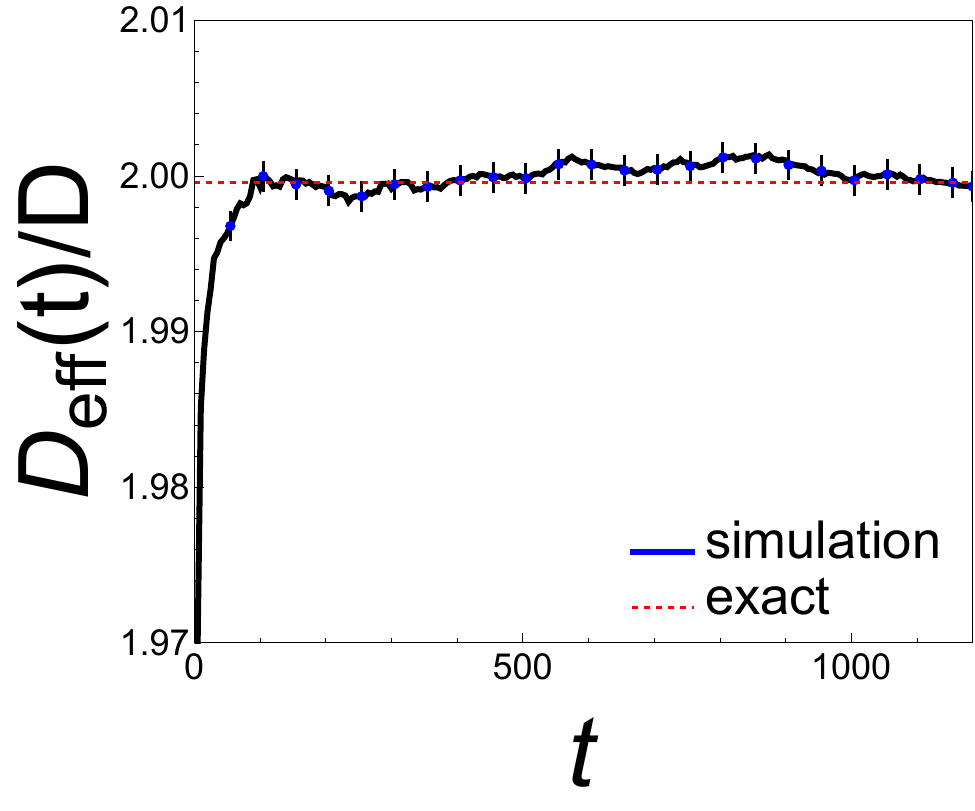}
\caption{
The normalized time-dependent diffusion coefficient \(D_{\rm eff}(t)/D\) for the driven system, calculated directly from particle trajectories.
Panels (a) and (b) show results obtained using \(10^6\) and \(10^7\) independent trajectories, respectively, with a time step
\(\Delta t=10^{-3}\).
The dashed red line represents the result obtained from Eq.~(\ref{eq:Deff-finite-driving}).
The horizontal axis shows the simulation time \(t\).
The parameters are \(L=v=\tau=1\), \(D=0.05\), and \(v_d=1.3\).
Error bars show the Gaussian estimate of the standard error.
}
\label{fig:Deff-vd-time}
\end{figure}

As a final test of Eq.~(\ref{eq:Deff-finite-driving}), Fig.~\ref{fig:Deff-vd-time} shows the time-dependent diffusion coefficient
\(D_{\rm eff}(t)\), calculated directly from the long-time displacement variance. As \(t\) increases, the trajectory results approach 
the value obtained from the stationary densities and currents of the physical and dual systems. The agreement provides an independent 
test of Eq.~(\ref{eq:Deff-finite-driving}) for a system with nonzero stationary current.

\section{Summary and conclusions}
\label{sec:summary}

This work develops a framework for calculating the effective diffusion coefficient of a Brownian particle in 
a one-dimensional fluctuating periodic force field. On spatial scales much larger than the period and at long times, the particle 
motion is characterized by an effective drift velocity \(v_{\rm eff}=Lj_{\rm ss}\) and an effective diffusion coefficient 
\(D_{\rm eff}\), which generally differs from the free-particle diffusion coefficient \(D\).

The framework is developed for a specific class of fluctuations. The force field \(F(x,\sigma)\) has the same spatial period \(L\) 
for every internal state \(\sigma\), while \(\sigma\) is independently redrawn from a fixed distribution \(p(\sigma)\) at 
Poisson-distributed times with rate \(\tau^{-1}\).

The framework is based on homogenization, which expresses \(D_{\rm eff}\) in terms of a corrector function satisfying a backward 
differential equation. Stochastic duality allows this equation to be reinterpreted as the forward equation of a system that is 
different from the physical system, which we refer to as the dual system. For the redraw dynamics considered here, the internal-state 
dynamics are reversible, and the dual system is obtained by reversing the force field, \(F\to-F\). 
A similar construction may be possible for more general reversible internal-state dynamics, although the resulting expression for 
\(D_{\rm eff}\) need not retain the form derived here.

For the class of fluctuations considered here, \(D_{\rm eff}\) can be expressed entirely in terms of the stationary densities 
and probability currents of the physical and dual systems:
\ba
D_{\rm eff}
&=&
DL^2\int_0^L dx\int d\sigma\,p(\sigma)n(x,\sigma)n_D^2(x,\sigma)
\nonumber\\
&&+
\frac{\tau L^2}{2}
\int_0^L dx\int d\sigma\,p(\sigma)n(x,\sigma)
\nonumber\\
&&\times
\int d\sigma'\,p(\sigma')
\Bigl[j_D(x,\sigma')-j_D(x,\sigma)\Bigr]^2 .
\label{eq:Deff-nd-summary}
\ea
The dual representation also reveals two symmetry relations:
\[
j_{\rm ss}[-F]=-j_{\rm ss}[F],
\qquad
D_{\rm eff}[-F]=D_{\rm eff}[F].
\]
Thus, reversing the force changes the direction of the stationary current but leaves the long-time diffusion coefficient unchanged.

The formula above can be repackaged by introducing the auxiliary function
\ba
Q(x)
&=&
D\int d\sigma\,p(\sigma)n(x,\sigma)n_D(x,\sigma)
\nonumber\\
&-&
\tau\int d\sigma\,p(\sigma)j(x,\sigma)j_D(x,\sigma).
\label{eq:Q-summary}
\ea
The effective diffusion coefficient can then be written as
\ba
D_{\rm eff}
&=&
L^2\int_0^L dx\,\bar n_D(x)Q(x)
\nonumber\\
&+&
\tau L^2j_{\rm ss}
\int_0^L dx\int d\sigma\,p(\sigma)n(x,\sigma)j_D(x,\sigma).
\label{eq:Deff-Q-summary}
\ea
The special property of \(Q(x)\) is that it obeys
\[
\partial_xQ(x)
=
j_{\rm ss}\bigl[\bar n(x)-\bar n_D(x)\bigr].
\]
Thus, when \(j_{\rm ss}=0\), not only does the second term in Eq.~(\ref{eq:Deff-Q-summary}) vanish, but \(Q(x)\) also becomes
independent of position, \(Q(x)=\mathrm{const}\). Using the normalization
\(
\int_0^L dx\,\bar n_D(x)=1,
\)
Eq.~(\ref{eq:Deff-Q-summary}) then reduces to
\[
D_{\rm eff}=L^2Q.
\]
Using the definition of \(Q\), this result can equivalently be written as
\[
D_{\rm eff}
=
L^2\int d\sigma\,p(\sigma)
\left[
D\,n(x,\sigma)n_D(x,\sigma)
-\tau\,j(x,\sigma)j_D(x,\sigma)
\right].
\]
In the absence of force-field fluctuations, this expression reduces to the classical Lifson--Jackson relation. We therefore refer
to its extension to fluctuating periodic force fields as the generalized Lifson--Jackson relation.

We tested the framework by applying it first to a non-fluctuating periodic force field. This provides a basic consistency check: 
the general expression reproduces the known result for diffusion in a tilted periodic potential and, when the driving force vanishes, 
reduces to the classical Lifson--Jackson formula.

We then considered a triangular potential with two types of fluctuations, \(\sigma\in\{-1,1\}\) and 
\(\sigma\in\{0,1\}\). Both models have zero stationary current and therefore allow us to apply the generalized 
Lifson--Jackson relation. The resulting analytical expressions for \(D_{\rm eff}\) were confirmed by direct simulations 
of the long-time particle displacement.

These examples also allowed us to examine the independent-cell first-passage construction,
\[
D_{\rm eff}^{\rm FP}(1)=\frac{L^2}{2\tau_f},
\]
where \(\tau_f\) is the mean time required to reach either neighboring cell. For both models, this construction 
overestimates the exact effective diffusion coefficient. The discrepancy arises from treating successive cell passages 
as independent lattice jumps. When the particle enters a new cell, however, it retains the force state inherited from 
the preceding passage and therefore carries memory of its previous motion.

The error can be reduced by extending the coarse-graining length to \(mL\), where \(m=1,2,\ldots\). The resulting estimate 
\(D_{\rm eff}^{\rm FP}(m)\) approaches the exact result as \(m\) increases. Although the independent-cell construction is valid 
for a non-fluctuating periodic potential, care is required when the force field fluctuates. For the fluctuations considered 
in this work, its single-cell form does not yield the correct effective diffusion coefficient.

Finally, we briefly considered a driven fluctuating system with \(j_{\rm ss}\neq0\), for which the generalized Lifson--Jackson 
relation does not apply. In this case, we used the general dual representation to calculate \(D_{\rm eff}\) from the stationary 
densities and currents of the physical and dual systems. The result agrees with both the numerical solution of the stationary 
equations and direct simulations of the long-time displacement variance. This example confirms that the general framework remains 
valid beyond the zero-current case and demonstrates its practical advantage: \(D_{\rm eff}\) can be obtained from stationary 
simulations within a single periodic cell, avoiding computationally expensive long-time trajectory simulations.

Although the dual representation derived here is specific to the Poisson redraw process, homogenization can be applied to broader 
classes of fluctuating periodic force fields. Whether the resulting corrector can be represented in terms of stationary quantities 
of a dual system, and what form that dual system takes, will depend on the dynamics of the fluctuations. Extending the present 
construction to other types of fluctuations, including nonreversible internal-state dynamics, remains an interesting direction 
for future work.

\begin{acknowledgments}
D.F. acknowledges financial support from FONDECYT through grant number 1241694.  
\end{acknowledgments}

\section{Data availability}
The data that support the findings of this study are available from the corresponding author upon 
reasonable request.


\appendix

\section{Establishing the relation \(\partial_xQ(x)=j_{\rm ss}[\bar n(x)-\bar n_D(x)]\)}
\label{app:dQ}

In this appendix, we establish the relation stated in Eq.~(\ref{eq:dQ}) and displayed in the section title.
The quantity \(Q(x)\), defined in Eq.~(\ref{eq:Q-continuous}), 
is reproduced below for convenience:
\[
\begin{aligned}
Q(x)
& = D\int d\sigma\,p(\sigma)n(x,\sigma)n_D(x,\sigma)\\
& \quad  -  \tau\int d\sigma\,p(\sigma)j(x,\sigma)j_D(x,\sigma).
\end{aligned}
\]

We begin by establishing two relations used in the derivation. The stationary forward equations 
of the original and dual systems can be written as
\begin{align}
\partial_x j(x,\sigma)
&= -\frac1\tau \left[ n(x,\sigma)-\bar n(x) \right], \nonumber\\
\partial_x j_D(x,\sigma)
&= -\frac1\tau \left[ n_D(x,\sigma)-\bar n_D(x) \right],
\label{eq:nice-app-dQ}
\end{align}
where 
\(
\bar n(x) \!\!=\!\! \int d\sigma p(\sigma)n(x,\sigma)
\) 
and 
\(
\bar n_D(x) \!\!=\!\! \int d\sigma p(\sigma)n_D(x,\sigma)
\)
are the marginal distributions. The corresponding probability currents are
\(
j=\mu Fn-D\partial_xn  
\)
and
\(
j_D = -\mu Fn_D - D\partial_xn_D.
\)
Multiplying the definition of \(j\) by \(n_D\), the definition of \(j_D\) by \(n\), and adding the resulting expressions gives
\be
n_Dj + n j_D = - D\partial_x(nn_D).
\label{eq:identity-app}
\ee
Eqs.~(\ref{eq:nice-app-dQ}) and (\ref{eq:identity-app}) constitute the two identities from which the relation in the title follows.

We now differentiate the definition of \(Q(x)\). For the first term, we use Eq.~(\ref{eq:identity-app}), 
multiplied by \(p(\sigma)\) and integrated over the force states. This gives
\begin{align}
\partial_xQ
&= -\int d\sigma\,p(\sigma) \left[ n_Dj+n j_D \right] \nonumber\\
&\qquad - \tau \int d\sigma\,p(\sigma) \left[ j_D\partial_xj + j\partial_xj_D \right].
\end{align}
Substituting the expressions for \(\partial_xj\) and \(\partial_xj_D\) from Eq.~(\ref{eq:nice-app-dQ}), we obtain
\begin{align}
\partial_xQ
&= - \int d\sigma\,p(\sigma) (n_Dj+n j_D) \nonumber\\
&\quad + \int d\sigma\,p(\sigma) \left[ j_D(n-\bar n) + j(n_D-\bar n_D) \right].
\end{align}
After canceling the terms that appear with opposite signs, the expression reduces to
\begin{align}
\partial_xQ
&= - \bar n \int d\sigma\,p(\sigma)j_D  - \bar n_D \int d\sigma\,p(\sigma)j.
\end{align}
Finally, the currents averaged over the force states satisfy
\(
\int d\sigma\,p(\sigma)j(x,\sigma) = j_{\rm ss}
\)
and 
\( 
\int d\sigma\,p(\sigma )j_D(x,\sigma) = -j_{\rm ss},
\)
where the second relation follows from Eq.~(\ref{eq:v-eff-dual-average}). Substituting these relations gives
\[
\partial_xQ(x) = j_{\rm ss}\left[\bar n(x)-\bar n_D(x)\right].
\]
This recovers Eq.~(\ref{eq:dQ}) and completes the proof.

\section{Effective diffusion in terms of \(Q(x)\)}
\label{app:Deff3}

In this appendix, we derive Eq.~(\ref{eq:Deff-3A}), which expresses the effective diffusion coefficient \(D_{\rm eff}\) 
in terms of \(Q(x)\).  The quantity $Q(x)$ has been defined in Eq.~(\ref{eq:Q-continuous}).  
We begin with the formula for \(D_{\rm eff}\) in Eq.~(\ref{eq:Dloc-dual-current-2}), reproduced below for convenience:
\ba
D_{\rm eff} &=& DL^2 \int_0^L dx \int d\sigma\,p(\sigma)\, n(x,\sigma)n_D^2(x,\sigma) \nonumber\\
&& + \frac{\tau L^2}{2} \int_0^L dx \int d\sigma\,p(\sigma)\, n(x,\sigma) \nonumber\\
&&\times \int d\sigma'\,p(\sigma') \Bigl[ j_D(x,\sigma')-j_D(x,\sigma) \Bigr]^2 .
\label{eq:Deff-start-app}
\ea
To make the derivation easier to follow, we treat the two contributions to \(D_{\rm eff}\) separately by defining
\be
A(x) = 
\int d\sigma\, p(\sigma)n(x,\sigma)
\int d\sigma' \, p(\sigma') \left[ j_D(x,\sigma') - j_D(x,\sigma) \right]^2
\label{eq:A-def}
\ee
and
\be
B(\sigma) = \int_0^L dx \, n(x,\sigma)n_D^2(x,\sigma).
\label{eq:B-def}
\ee
In terms of these two quantities, the effective diffusion coefficient becomes
\be
D_{\rm eff}  = DL^2 \int d\sigma \, p(\sigma)B(\sigma) + \frac{\tau L^2}{2}\int_0^L dx \,A(x).
\label{eq:Deff-AB}
\ee

We start by expanding \(A(x)\) 
\be
A(x) = \bar n\int d\sigma\,p(\sigma)j_D^2 + \int d\sigma\,p(\sigma)n j_D^2 + 2 j_{\rm ss} \int d\sigma\,p(\sigma)n j_D
\label{eq:A-fin}
\ee
where \( \bar n(x) = \int d\sigma\,p(\sigma)n(x,\sigma)\) is the marginal density, already defined in Appendix~\ref{app:dQ}.  
We also used 
\(
j_{\rm ss} = -\int d\sigma\,p(\sigma)j_D(x,\sigma).  
\)

Next, we rewrite \(B(\sigma)\). Using Eq.~(\ref{eq:nice-app-dQ}), one of the two factors of \(n_D\) can be expressed as
\be
n_D(x,\sigma) = \bar n_D(x)-\tau\partial_xj_D(x,\sigma).
\label{eq:nD-current-app}
\ee
Substituting this relation into the definition of \(B(\sigma)\) and integrating the resulting second term by parts, 
then using periodic boundary conditions, gives
\be
B(\sigma) = \int_0^L dx\,\bar n_D\,n\,n_D + \tau\int_0^L dx\,j_D\,\partial_x(nn_D).
\ee
Using another identity derived in the previous appendix, Eq.~(\ref{eq:identity-app}), we can write
\(
-D\partial_x(nn_D) = n_Dj + n j_D.  
\)
Substituting this relation into the expression for \(B(\sigma)\) gives 
\be
B(\sigma) =  \int_0^L dx \,\bar n_D  \, n \, n_D    -    \frac{\tau}{D} \int_0^L dx \, n_D\, j_D j     -    \frac{\tau}{D} \int_0^L dx \, n\, j_D^2. 
\label{eq:B-fin}
\ee

Substituting the expressions for \(A(x)\) and \(B(\sigma)\) from Eqs.~(\ref{eq:A-fin}) and (\ref{eq:B-fin}) into Eq.~(\ref{eq:Deff-AB}), 
and collecting the resulting terms, gives
\ba
\frac{D_{\rm eff}}{L^2}
&=& D\int_0^L dx\,\bar n_D \int d\sigma\,p n n_D \nonumber\\
&-&
\tau\int_0^L dx\int d\sigma\,p n_Dj j_D \nonumber\\
&-&
\frac{\tau}{2}\int_0^L dx\int d\sigma\, p n j_D^2 \nonumber\\
&+&
 \frac{\tau}{2}\int_0^L dx\,\bar n \int d\sigma\,p j_D^2 \nonumber\\
&+&
 \tau j_{\rm ss}\int_0^L dx\int d\sigma\,p n j_D .
\label{eq:Deff-app-middle}
\ea

We now use the definition
\[
Q(x)  =  D\int d\sigma\,p(\sigma)\,nn_D   -   \tau\int d\sigma\,p(\sigma)\,j j_D
\]
to reorganize Eq.~(\ref{eq:Deff-app-middle}). This gives
\be
\frac{D_{\rm eff}}{L^2} = \int_0^L dx\,\bar n_D(x)Q(x)   +   \tau j_{\rm ss} \int_0^L dx\int d\sigma\,p(\sigma)\,n j_D + \tau R,
\label{eq:Deff-app-R}
\ee
where \(R\) collects the terms that have not yet been incorporated into \(Q(x)\):
\ba
R &=& \int_0^L dx\int d\sigma\,p(\sigma)\,j j_D(\bar n_D-n_D) \nonumber\\
&&-\frac12\int_0^L dx\int d\sigma\,p(\sigma)\,(n-\bar n)j_D^2.
\label{eq:R-app}
\ea
It remains to show that \(R=0\).

We use the relations derived in Eq.~(\ref{eq:nice-app-dQ}), repeated below for convenience:
\[
\bar n_D-n_D=\tau\partial_xj_D,
\qquad
n-\bar n=-\tau\partial_xj.
\]
Substituting them into Eq.~(\ref{eq:R-app}) gives
\[
R = \tau\int_0^L dx\int d\sigma\,p(\sigma)\, j j_D\partial_xj_D + \frac{\tau}{2} \int_0^L dx\int d\sigma\,p(\sigma)\, (\partial_xj)j_D^2.
\]
The two terms combine into a total derivative,
\[
R
=
\frac{\tau}{2}
\int_0^L dx\int d\sigma\,p(\sigma)\,
\partial_x\!\left(jj_D^2\right).
\]
Because \(j\) and \(j_D\) are periodic in \(x\), this integral vanishes, and therefore \(R=0\). 
Substituting this result into Eq.~(\ref{eq:Deff-app-R}) gives
\be
\frac{D_{\rm eff}[F]}{L^2} = \int_0^L dx\,\bar n_D(x)Q(x) + \tau j_{\rm ss} \int_0^L dx\int d\sigma\,p(\sigma)\,n j_D.
\label{eq:Deff-app-Qdual}
\ee

\section{Numerical calculation of the mean first-passage time}
\label{app:MFPT-m}

In this appendix, we describe the numerical calculation of the mean first-passage time \(\tau_f^{(m)}\) used 
in the first-passage estimate
\be
D_{\rm eff}^{\rm FP}(m) = \frac{m^2L^2}{2\tau_f^{(m)}}.
\label{eq:Deff-FP-m-app}
\ee
The particle starts at \(x=0\), and the first-passage event occurs when it reaches either \(x=-mL\) or \(x=mL\). 
Thus, \(mL\) is the distance associated with a coarse-grained jump.

\subsection{The two-state model \(\sigma\in\{-1,1\}\)}

We first consider the model in which the force switches between \(F(x)\) and \(-F(x)\). Let \(T_+^{(m)}(x)\) and \(T_-^{(m)}(x)\) 
denote the mean first-passage times for a particle initially at position \(x\) in states \(+\) and \(-\), respectively. 
These quantities satisfy the coupled backward equations
\begin{align}
-1
&=
\mu F(x)\partial_xT_+^{(m)}
+
D\partial_x^2T_+^{(m)}
+
\frac{1}{2\tau}
\left(
T_-^{(m)}-T_+^{(m)}
\right),
\label{eq:MFPT-plus-app}
\\
-1
&=
-\mu F(x)\partial_xT_-^{(m)}
+
D\partial_x^2T_-^{(m)}
+
\frac{1}{2\tau}
\left(
T_+^{(m)}-T_-^{(m)}
\right).
\label{eq:MFPT-minus-app}
\end{align}
Redraw events occur at rate \(1/\tau\), and the new state is selected as \(+\) or \(-\) with equal probability. 
The transition rate from either state to the other is therefore \(1/(2\tau)\).
The absorbing boundaries at \(x=\pm mL\) impose
\be
T_+^{(m)}(-mL)
=
T_-^{(m)}(-mL)
=
T_+^{(m)}(mL)
=
T_-^{(m)}(mL)
=
0.
\label{eq:MFPT-boundary-app}
\ee
The mean first-passage times and their first derivatives are continuous, even across the discontinuity of force
\[
T_\pm^{(m)}(x^-)=T_\pm^{(m)}(x^+),
\qquad
\partial_xT_\pm^{(m)}(x^-)=\partial_xT_\pm^{(m)}(x^+).
\]
Their second derivatives may be discontinuous. These matching conditions follow by integrating the backward 
equations across an infinitesimal interval containing the force discontinuity.

Because the initial force state is selected as \(+\) or \(-\) with equal probability, the mean first-passage time entering 
Eq.~(\ref{eq:Deff-FP-m-app}) is
\be
\tau_f^{(m)} = \frac{T_+^{(m)}(0)+T_-^{(m)}(0)}{2}.
\label{eq:MFPT-average-app}
\ee

We solve Eqs.~(\ref{eq:MFPT-plus-app}) and (\ref{eq:MFPT-minus-app}) numerically using finite differences. 
The interval \([-mL,mL]\) is divided into a uniform grid,
\[
x_k = -mL + kh,
\qquad
k=0,\ldots,2mN_L,
\qquad
h=\frac{L}{N_L},
\]
where \(N_L\) is the number of grid intervals per spatial period. We choose \(N_L\) to be even so that \(x=0\), 
together with all discontinuities of the triangular force, lies on the grid.

For each interior grid point, \(k=1,\ldots,2mN_L-1\), the derivatives are approximated by
\[
\partial_xT(x_k)
\simeq
\frac{T_{k+1}-T_{k-1}}{2h},
\qquad
\partial_x^2T(x_k)
\simeq
\frac{T_{k+1}-2T_k+T_{k-1}}{h^2}.
\]
The discretized backward equations are
\begin{align}
-1
&=
\mu F(x_k)
\frac{
T_{+,k+1}^{(m)}-T_{+,k-1}^{(m)}
}{2h}
+
D\frac{
T_{+,k+1}^{(m)}-2T_{+,k}^{(m)}+T_{+,k-1}^{(m)}
}{h^2}
\nonumber\\
&\quad
+
\frac{
T_{-,k}^{(m)}-T_{+,k}^{(m)}
}{2\tau},
\label{eq:MFPT-plus-discrete-app}
\\
-1
&=
-\mu F(x_k)
\frac{
T_{-,k+1}^{(m)}-T_{-,k-1}^{(m)}
}{2h}
+
D\frac{
T_{-,k+1}^{(m)}-2T_{-,k}^{(m)}+T_{-,k-1}^{(m)}
}{h^2}
\nonumber\\
&\quad
+
\frac{
T_{+,k}^{(m)}-T_{-,k}^{(m)}
}{2\tau}.
\label{eq:MFPT-minus-discrete-app}
\end{align}
Together with the absorbing boundary conditions in Eq.~(\ref{eq:MFPT-boundary-app}), these equations 
form a sparse linear system for \(T_{+,k}^{(m)}\) and \(T_{-,k}^{(m)}\) at the interior grid points.

After solving this system, we evaluate \(T_+^{(m)}\) and \(T_-^{(m)}\) at \(x=0\), calculate \(\tau_f^{(m)}\) from 
Eq.~(\ref{eq:MFPT-average-app}), and substitute the result into Eq.~(\ref{eq:Deff-FP-m-app}). Repeating this 
calculation for different values of \(m\) gives the first-passage estimate as a function of the coarse-graining length.

The calculations presented in the main text were performed using \(N_L=2000\) 
grid intervals per spatial period. Numerical convergence was verified by repeating the calculations with 
\(N_L=500\) and \(N_L=1000\).

\subsection{The two-state model \(\sigma\in\{0,1\}\)}

For the on--off model, we define \(T_{\rm on}^{(m)}(x)\) and \(T_{\rm off}^{(m)}(x)\) as the mean first-passage times in the two states. 
The corresponding backward equations are
\begin{align}
-1
&=
\mu F(x)\partial_xT_{\rm on}^{(m)}
+
D\partial_x^2T_{\rm on}^{(m)}
+
\frac{T_{\rm off}^{(m)}-T_{\rm on}^{(m)}}{2\tau},
\\
-1
&=
D\partial_x^2T_{\rm off}^{(m)}
+
\frac{T_{\rm on}^{(m)}-T_{\rm off}^{(m)}}{2\tau}.
\end{align}
These equations are subject to absorbing boundary conditions at \(x=\pm mL\) and are solved using the finite-difference 
procedure described above. We initialize the two states with equal probability and define
\[
\tau_f^{(m)} =
\frac{T_{\rm on}^{(m)}(0)+T_{\rm off}^{(m)}(0)}{2}.
\]
Substitution into Eq.~(\ref{eq:Deff-FP-m-app}) then gives \(D_{\rm eff}^{\rm FP}(m)\).

\bibliography{general}

@book{book-Klafter-2011,
  author    = {Klafter, J. and Sokolov, I. M.},
  title     = {First Steps in Random Walks: From Tools to Applications},
  publisher = {Oxford University Press},
  address   = {Oxford},
  year      = {2011},
  doi       = {10.1093/acprof:oso/9780199234868.001.0001}
}

@article{JMP-Montroll-1965,
  author  = {Montroll, Elliott W. and Weiss, George H.},
  title   = {Random Walks on Lattices. II},
  journal = {J. Math. Phys.},
  volume  = {6},
  pages   = {167--181},
  year    = {1965},
  doi     = {10.1063/1.1704269}
}

@article{JCP-Frydel-2026,
  author  = {Frydel, Derek},
  title   = {Forward--backward correspondence between stationary structure and splitting probabilities in active matter},
  journal = {The Journal of Chemical Physics},
  volume  = {165},
  number  = {16},
  pages   = {164902},
  year    = {2026}
}

@article{EPL-Schehr-2020,
  author  = {Pierre {Le Doussal} and Satya N. Majumdar and Gr{\'e}gory Schehr},
  title   = {Velocity and diffusion constant of an active particle in a one-dimensional force field},
  journal = {EPL},
  volume  = {130},
  number  = {4},
  pages   = {40002},
  year    = {2020},
  doi     = {10.1209/0295-5075/130/40002}
}

@article{PD-Vergassola-1997,
  author  = {Vergassola, Massimo and Avellaneda, Marco},
  title   = {Scalar transport in compressible flow},
  journal = {Physica D: Nonlinear Phenomena},
  volume  = {106},
  number  = {1-2},
  pages   = {148--166},
  year    = {1997},
  doi     = {10.1016/S0167-2789(97)00022-5}
}

@article{AP-Bhattacharya-1985,
  author  = {Rabi Bhattacharya},
  title   = {A Central Limit Theorem for Diffusions with Periodic Coefficients},
  journal = {The Annals of Probability},
  volume  = {13},
  number  = {2},
  pages   = {385--396},
  year    = {1985},
  doi     = {10.1214/aop/1176992998}
}

@article{PRE-Frydel-2026,
  title = {Semi-Markovian switching in a fluctuating harmonic trap: An age-structured formulation},
  author = {Frydel, Derek},
  journal = {Phys. Rev. E},
  volume = {114},
  issue = {2},
  pages = {024105},
  numpages = {13},
  year = {2026},
  month = {Aug},
  publisher = {American Physical Society},
  doi = {10.1103/7pxv-tg3b},
  url = {https://link.aps.org/doi/10.1103/7pxv-tg3b}
}

@article{AP-Abad-1998,
  author  = {Abad, Enrique and Mielke, Andreas},
  title   = {Brownian motion in fluctuating periodic potentials},
  journal = {Annalen der Physik},
  volume  = {7},
  number  = {1},
  pages   = {9--23},
  year    = {1998}
}

@article{PRL-Astumian-1994,
  author  = {Astumian, R. Dean and Bier, Martin},
  title   = {Fluctuation driven ratchets: Molecular motors},
  journal = {Physical Review Letters},
  volume  = {72},
  pages   = {1766--1769},
  year    = {1994},
  doi     = {10.1103/PhysRevLett.72.1766}
}

@article{PRL-Faucheux-1995,
  author  = {Faucheux, L. P. and Bourdieu, L. S. and
             Kaplan, P. D. and Libchaber, A. J.},
  title   = {Optical thermal ratchet},
  journal = {Physical Review Letters},
  volume  = {74},
  pages   = {1504--1507},
  year    = {1995},
  doi     = {10.1103/PhysRevLett.74.1504}
}

@article{PR-Reimann-2002,
  author  = {Reimann, Peter},
  title   = {Brownian motors: Noisy transport far from equilibrium},
  journal = {Physics Reports},
  volume  = {361},
  pages   = {57--265},
  year    = {2002},
  doi     = {10.1016/S0370-1573(01)00081-3}
}

@article{RMP-Hanggi-2009,
  author  = {H{\"a}nggi, Peter and Marchesoni, Fabio},
  title   = {Artificial Brownian motors: Controlling transport on the nanoscale},
  journal = {Reviews of Modern Physics},
  volume  = {81},
  pages   = {387--442},
  year    = {2009},
  doi     = {10.1103/RevModPhys.81.387}
}

@article{PA-Gleeson-2009,
  author  = {Gleeson, James P.},
  title   = {Transport in randomly-fluctuating spatially-periodic potentials},
  journal = {Physica A},
  volume  = {388},
  pages   = {277--287},
  year    = {2009},
  doi     = {10.1016/j.physa.2008.10.013}
}

@article{EPJE-Su-2023,
  author  = {Su, Meng and Lindner, Benjamin},
  title   = {Active Brownian particles in a biased periodic potential},
  journal = {European Physical Journal E},
  volume  = {46},
  pages   = {22},
  year    = {2023},
  doi     = {10.1140/epje/s10189-023-00283-w}
}

@article{PRE-Roberts-2023,
  author  = {Roberts, Connor and Zhen, Zigan},
  title   = {Run-and-tumble motion in a linear ratchet potential:
             Analytic solution, power extraction, and first-passage properties},
  journal = {Physical Review E},
  volume  = {108},
  pages   = {014139},
  year    = {2023},
  doi     = {10.1103/PhysRevE.108.014139}
}

@article{SM-Koumakis-2014,
  author  = {Koumakis, Nikolaos and Maggi, Claudio and Di Leonardo, Roberto},
  title   = {Directed transport of active particles over asymmetric energy barriers},
  journal = {Soft Matter},
  volume  = {10},
  pages   = {5695--5701},
  year    = {2014},
  doi     = {10.1039/C4SM00665H}
}

@article{PRL-Lindner-2008,
  author  = {Lindner, Benjamin and Nicola, Ernesto M.},
  title   = {Critical asymmetry for giant diffusion of active Brownian particles},
  journal = {Physical Review Letters},
  volume  = {101},
  pages   = {190603},
  year    = {2008},
  doi     = {10.1103/PhysRevLett.101.190603}
}

@article{PRE-Akimoto-2025,
  author    = {Iida, Kento and Akimoto, Takuma and Dechant, Andreas},
  title     = {Universality of giant diffusion in tilted periodic potentials},
  journal   = {Physical Review E},
  volume    = {111},
  number    = {2},
  pages     = {024113},
  year      = {2025},
  publisher = {American Physical Society},
  doi       = {10.1103/PhysRevE.111.024113}
}

@article{PRE-Dubkov-2005,
  title = {Acceleration of diffusion in randomly switching potential with supersymmetry},
  author = {Dubkov, Alexander A. and Spagnolo, Bernardo},
  journal = {Phys. Rev. E},
  volume = {72},
  issue = {4},
  pages = {041104},
  numpages = {8},
  year = {2005},
  month = {Oct},
  publisher = {American Physical Society},
  doi = {10.1103/PhysRevE.72.041104},
  url = {https://link.aps.org/doi/10.1103/PhysRevE.72.041104}
}

@article{EPJB-Dubkov-2006,
  author  = {Dubkov, A. A. and Spagnolo, B.},
  title   = {Diffusion in flashing periodic potentials},
  journal = {The European Physical Journal B},
  volume  = {50},
  number  = {3},
  pages   = {299--303},
  year    = {2006},
  doi     = {10.1140/epjb/e2006-00108-y}
}

@article{EPJE-Calisto-2017,
  author    = {Calisto, Hector and Bologna, Mauro and Chand{\'i}a, Kristopher J.},
  title     = {Forced dichotomic diffusion in a viscous media},
  journal   = {The European Physical Journal B},
  year      = {2017},
  volume    = {90},
  number    = {2},
  pages     = {24},
  doi       = {10.1140/epjb/e2016-70643-y},
  url       = {https://doi.org/10.1140/epjb/e2016-70643-y}
}

@article{AP-Siegmund-1976,
  author    = {Siegmund, D.},
  title     = {The Equivalence of Absorbing and Reflecting Barrier Problems for Stochastically Monotone {M}arkov Processes},
  journal   = {The Annals of Probability},
  volume    = {4},
  number    = {6},
  pages     = {914--924},
  month     = {December},
  year      = {1976},
  publisher = {Institute of Mathematical Statistics},
  doi       = {10.1214/aop/1176995936},
  url       = {https://doi.org/10.1214/aop/1176995936}
}

@article{JPA-Gueneau-2024,
doi = {10.1088/1751-8121/ad4753},
url = {https://doi.org/10.1088/1751-8121/ad4753},
year = {2024},
month = {may},
publisher = {IOP Publishing},
volume = {57},
number = {22},
pages = {225005},
author = {Guéneau, Mathis and Touzo, Léo},
title = {Relating absorbing and hard wall boundary conditions for a one-dimensional run-and-tumble particle},
journal = {Journal of Physics A: Mathematical and Theoretical}
}

@article{JSTAT-Gueneau-2024,
doi = {10.1088/1742-5468/ad6134},
url = {https://doi.org/10.1088/1742-5468/ad6134},
year = {2024},
month = {sep},
publisher = {IOP Publishing},
volume = {2024},
number = {8},
pages = {083208},
author = {Guéneau, Mathis and Touzo, Léo},
title = {Siegmund duality for physicists: a bridge between spatial and first-passage properties of continuous- and discrete-time stochastic processes},
journal = {Journal of Statistical Mechanics: Theory and Experiment}
}

@article{PRL-Mathis-2026,
  title = {Active Brownian Dynamics in Channels: First-Passage and Spatiotemporal Properties via Siegmund Duality},
  author = {Baouche, Yanis and Gu\'eneau, Mathis and Kurzthaler, Christina},
  journal = {Phys. Rev. Lett.},
  volume = {137},
  issue = {13},
  pages = {138301},
  numpages = {10},
  year = {2026},
  month = {Sep},
  publisher = {American Physical Society},
  doi = {10.1103/g83n-r4hs},
  url = {https://link.aps.org/doi/10.1103/g83n-r4hs}
}

@article{PRL-Reimann-2001,
  title = {Giant Acceleration of Free Diffusion by Use of Tilted Periodic Potentials},
  author = {Reimann, P. and Van den Broeck, C. and Linke, H. and H\"anggi, P. and Rubi, J. M. and P\'erez-Madrid, A.},
  journal = {Phys. Rev. Lett.},
  volume = {87},
  issue = {1},
  pages = {010602},
  numpages = {4},
  year = {2001},
  month = {Jun},
  publisher = {American Physical Society},
  doi = {10.1103/PhysRevLett.87.010602},
  url = {https://link.aps.org/doi/10.1103/PhysRevLett.87.010602}
}

@article{PRE-Reimann-2002,
  author  = {Reimann, P. and Van den Broeck, C. and Linke, H. and H{\"a}nggi, P. and Rub{\'\i}, J. M. and P{\'e}rez-Madrid, A.},
  title   = {Diffusion in Tilted Periodic Potentials: Enhancement, Universality, and Scaling},
  journal = {Physical Review E},
  volume  = {65},
  number  = {3},
  pages   = {031104},
  year    = {2002},
  doi     = {10.1103/PhysRevE.65.031104}
}

@article{JCP-Lifson-1962,
  author  = {Lifson, S. and Jackson, J. L.},
  title   = {On the Self-Diffusion of Ions in a Polyelectrolyte Solution},
  journal = {The Journal of Chemical Physics},
  volume  = {36},
  number  = {9},
  pages   = {2410--2414},
  year    = {1962},
  doi     = {10.1063/1.1732899}
}

@article{PRL-Reimann-2001a,
  title = {Supersymmetric Ratchets},
  author = {Reimann, Peter},
  journal = {Phys. Rev. Lett.},
  volume = {86},
  issue = {22},
  pages = {4992--4995},
  numpages = {0},
  year = {2001},
  month = {May},
  publisher = {American Physical Society},
  doi = {10.1103/PhysRevLett.86.4992},
  url = {https://link.aps.org/doi/10.1103/PhysRevLett.86.4992}
}

@book{book-Bensoussan-1978,
  title     = {Asymptotic Analysis for Periodic Structures},
  author    = {Bensoussan, Alain and Lions, Jacques-Louis and Papanicolaou, George},
  volume    = {5},
  year      = {1978},
  publisher = {North-Holland Publishing Company},
  address   = {Amsterdam}
}

@book{book-Pavliotis-2008,
  title     = {Multiscale Methods: Averaging and Homogenization},
  author    = {Pavliotis, Grigorios A. and Stuart, Andrew M.},
  volume    = {57},
  year      = {2008},
  publisher = {Springer Science \& Business Media},
  address   = {New York}
}

@article{JSP-Latorre-2013,
  author  = {Latorre, J. C. and Pavliotis, G. A. and Kramer, P. R.},
  title   = {Corrections to Einstein's Relation for Brownian Motion in a Tilted Periodic Potential},
  journal = {Journal of Statistical Physics},
  volume  = {150},
  pages   = {776--803},
  year    = {2013},
  doi     = {10.1007/s10955-013-0692-1}
}

@article{FNL-Pavliotis-2008,
  author  = {Pavliotis, G. A. and Vogiannou, A.},
  title   = {Diffusive Transport in Periodic Potentials: Underdamped Dynamics},
  journal = {Fluctuation and Noise Letters},
  volume  = {8},
  number  = {2},
  pages   = {L155--L173},
  year    = {2008},
  doi     = {10.1142/S0219477508004453}
}

@book{book-Watanabe-1989,
  author    = {Ikeda, Nobuyuki and Watanabe, Shinzo},
  title     = {Stochastic Differential Equations and Diffusion Processes},
  edition   = {2},
  publisher = {North-Holland},
  year      = {1989}
}

@book{book-Yin-Zhu-2010,
  author    = {Yin, G. George and Zhu, Chao},
  title     = {Hybrid Switching Diffusions: Properties and Applications},
  publisher = {Springer},
  year      = {2010},
  doi       = {10.1007/978-1-4419-1105-6}
}

@article{Lindner-FNL-2001,
  title   = {Optimal diffusive transport in a tilted periodic potential},
  author  = {Lindner, Benjamin and Kostur, Marcin and Schimansky-Geier, Lutz},
  journal = {Fluctuation and Noise Letters},
  volume  = {1},
  number  = {1},
  pages   = {R25--R39},
  year    = {2001},
  doi     = {10.1142/S0219477501000056}
}

@article{PRE-Frydel-2022,
  title = {Intuitive view of entropy production of ideal run-and-tumble particles},
  author = {Frydel, Derek},
  journal = {Phys. Rev. E},
  volume = {105},
  issue = {3},
  pages = {034113},
  numpages = {8},
  year = {2022},
  month = {Mar},
  publisher = {American Physical Society},
  doi = {10.1103/PhysRevE.105.034113},
  url = {https://link.aps.org/doi/10.1103/PhysRevE.105.034113}
}

@article{PRE-Razin-2020,
  title = {Entropy production of an active particle in a box},
  author = {Razin, Nitzan},
  journal = {Phys. Rev. E},
  volume = {102},
  issue = {3},
  pages = {030103},
  numpages = {6},
  year = {2020},
  month = {Sep},
  publisher = {American Physical Society},
  doi = {10.1103/PhysRevE.102.030103},
  url = {https://link.aps.org/doi/10.1103/PhysRevE.102.030103}
}

@article{POF-Frydel-2024,
    author = {Frydel, Derek},
    title = "{Run-and-tumble particles in slit geometry as a splitting probability problem}",
    journal = {Physics of Fluids},
    volume = {36},
    number = {11},
    pages = {111901},
    year = {2024},
    month = {11},
    doi = {10.1063/5.0233462},
    url = {https://doi.org/10.1063/5.0233462},
}

@article{JSTAT-Frydel-2021,
doi = {10.1088/1742-5468/ac1665},
url = {https://dx.doi.org/10.1088/1742-5468/ac1665},
year = {2021},
month = {aug},
publisher = {IOP Publishing and SISSA},
volume = {2021},
number = {8},
pages = {083220},
author = {Derek Frydel},
title = {Generalized run-and-tumble model in 1D geometry for an arbitrary distribution of drift velocities},
journal = {Journal of Statistical Mechanics: Theory and Experiment}
}

\end{document}